\documentclass[pra, aps,letterpaper, preprintnumbers,superscriptaddress]{revtex4}
\usepackage{hyperref}
\usepackage{verbatim}
\usepackage{amsmath}
\usepackage{latexsym}
\usepackage{revsymb}
\usepackage{yfonts}
\usepackage{ifthen}

\usepackage{natbib}
\usepackage{amsfonts}
\usepackage{amsmath}
\usepackage{amssymb}
\usepackage{amsthm}
\usepackage{graphicx}
\usepackage{bm}
\usepackage{bbm}

\newcommand{\be}{\begin{eqnarray} \begin{aligned}}
\newcommand{\ee}{\end{aligned} \end{eqnarray} }
\newcommand{\benn}{\begin{eqnarray*} \begin{aligned}}
\newcommand{\eenn}{\end{aligned} \end{eqnarray*} }

\newcommand{\bc}{\begin{center}}
\newcommand{\ec}{\end{center}}

\newcommand{\epscor}{\epsilon_{\textrm{cor}}}
\newcommand{\epssec}{\epsilon_{\textrm{sec}}}
\newcommand{\tn}[1]{\textnormal{#1}}

\newcommand{\HmaxOp}{H_{\max}}
\newcommand{\HminOp}{H_{\min}}

\usepackage{amsfonts}

\def\01{\{0,1\}}

\newcommand{\eps}{\varepsilon}
\newcommand{\ket}[1]{|#1\rangle}

\newcommand{\proj}[1]{|#1\rangle\langle#1|}

\newcommand{\assign}{\ensuremath{\kern.5ex\raisebox{.1ex}{\mbox{\rm:}}\kern -.3em =}}

\newcounter{protoCount}
\newcounter{protoList}
\newsavebox{\tmpbox}
\newlength{\protobox}

\begin{document}

\title{Finite-key analysis for measurement-device-independent quantum key distribution}

\author{Marcos Curty}
\affiliation{EI Telecomunicaci\'on, Dept. of Signal Theory and Communications, University of Vigo, E-36310 Vigo, Spain}
\author{Feihu Xu}
\affiliation{Center for Quantum Information and Quantum Control, Dept. of Physics and Dept. of 
Electrical \& Computer Engineering, University of Toronto, M5S 3G4 Toronto, Canada}
\author{Wei Cui}
\affiliation{Center for Quantum Information and Quantum Control, Dept. of Physics and Dept. of 
Electrical \& Computer Engineering, University of Toronto, M5S 3G4 Toronto, Canada}
\author{Charles Ci Wen Lim}
\affiliation{Group of Applied Physics, University of Geneva, CH-1211 Geneva, Switzerland}
\author{Kiyoshi Tamaki}
\affiliation{NTT Basic Research Laboratories, NTT Corporation, 3-1, Morinosato Wakamiya Atsugi-Shi,
Kanagawa, 243-0198, Japan}
\author{Hoi-Kwong Lo}
\affiliation{Center for Quantum Information and Quantum Control, Dept. of Physics and Dept. of 
Electrical \& Computer Engineering, University of Toronto, M5S 3G4 Toronto, Canada}

\begin{abstract}
Quantum key distribution promises unconditionally secure communications. 
However, as practical devices tend to deviate from their specifications, the security of some practical systems is no longer valid. In particular, 
an adversary can exploit imperfect detectors to learn a large part of the secret key, even though the security proof claims otherwise. 
Recently, a practical approach---measurement-device-independent quantum key distribution---has been  
proposed to solve this problem.
However, so far its security 
has only been fully proven under the assumption that the legitimate users of the system have  
unlimited resources. Here we fill this gap and provide a rigorous security proof against general attacks in the finite-key regime.
This is obtained by applying large deviation
theory, specifically the Chernoff bound, to perform parameter estimation. 
For the first time we demonstrate  
the feasibility of long-distance
implementations of measurement-device-independent quantum key distribution within a reasonable time-frame of signal transmission.  
\end{abstract}
\maketitle

\section{Introduction}

It is unequivocal that quantum key distribution (QKD)~\cite{QKDreview1,QKDreview2} needs to bridge the gap between theory and practice. In theory, 
QKD offers perfect security. In practice, 
however, 
it does not, as most practical devices behave differently from the theoretical models assumed in the security proofs. 
As a result, we face 
implementation loopholes, or so-called side-channels, which may be used by adversaries without being detected,
as seen
in recent attacks against certain commercial QKD 
systems~\cite{sidechannelrefs2a, sidechannelrefs1a, sidechannelrefs2b, sidechannelrefs2c, sidechannelrefs1b, sidechannelrefs1c, sidechannelrefs2d, sidechannelrefs2f, sidechannelrefs2e}. 

There are two potential ways to 
guarantee security in the  
realisations of QKD. The first is to 
develop mathematical models that perfectly match the behaviour of physical apparatuses, 
and then incorporate this information into a new security proof. 
While this is plausible in theory, unfortunately it is very hard to realise in practice, if not impossible.  
The second alternative is to design new protocols and develop security proof techniques 
that are 
compatible with a wide class of device 
imperfections. 
This allows us to omit an
accurate characterisation of real apparatuses.
The most well-known example of such a solution is 
(full) device-independent QKD (diQKD)~\cite{diqkdrefsa,diqkdrefsb,diqkdrefsc,diqkdrefsd,diqkdrefse}. Here, the legitimate users of the system 
(typically called Alice and Bob) treat their devices as two quasi ``black boxes", i.e., 
they need to know which elements their boxes contain, but not 
how they fully function~\cite{diqkdrefs2}. 
The security of diQKD relies 
on the violation of a Bell inequality~\cite{bell1,bell2}, which certifies the presence of quantum correlations. 
Despite its beauty, however,
this approach is highly impractical because it
requires a loophole-free Bell test which at the moment is still unavailable~\cite{Pearle1970}. 
Also, its secret key rate at practical distances is very limited~\cite{Gisin2010a,Gisin2010b}.

Very recently, 
a novel approach has been 
introduced, which is fully practical and feasible to implement. 
This scheme is known as 
measurement-device-independent QKD (mdiQKD)~\cite{mdiQKD} and 
offers a clear avenue to bridge
the gap between theory and practice. 
Its feasibility has been 
promptly demonstrated both in laboratories and 
via field-tests~\cite{Rubenok2012a,Rubenok2012b,Rubenok2012c,Rubenok2013a,Zhiyuan:2013}.
It successfully
removes
all (existing and yet to be discovered) detector side-channels~\cite{sidechannelrefs2a,sidechannelrefs2b,sidechannelrefs2c,sidechannelrefs2d,sidechannelrefs2e,sidechannelrefs2f}, 
which, arguably, is the most critical part of most QKD implementations.
Importantly, 
in contrast to diQKD this solution does not require that Alice and Bob perform a loophole-free Bell test; 
it is enough if they prove the presence of entanglement in a quantum state that is effectively distributed between them, 
just like in standard QKD schemes~\cite{curty_prl2004}.
In addition, now Alice and Bob may treat the measurement apparatus
as a true ``black box", which may be fully controlled 
by the adversary. A slight drawback is that
Alice and Bob need to characterise the quantum states 
[e.g., the polarisation degrees of freedom of phase-randomised weak coherent pulses (WCPs)]
that they send through 
the channel. But, as this process can be verified in a protected environment outside the influence of the 
adversary, it is less likely to be a problem. 
For completeness, the readers can refer to~\cite{charles2} where a characterisation of the prepared states is no longer required.

Nevertheless, so far the security of mdiQKD has only been proven 
in the asymptotic regime~\cite{mdiQKD}, which assumes that Alice and Bob have access to an unlimited amount of resources,
or in the finite regime but only against particular types of attacks~\cite{finitea,finiteb}.
In summary, until now, a rigorous security proof of mdiQKD that takes full 
account of the finite size effects~\cite{finite1,finite2,finite3} has appeared to be missing and, for this reason, the 
feasibility of long-distance implementations of mdiQKD within a reasonable time-frame of signal transmission 
has remained undemonstrated.

The main contributions of this work are twofold. First, in contrast to existing heuristic results on 
mdiQKD, we provide, for the first time, a security proof in the finite-key regime 
that is valid against general attacks, and
satisfies the composability definition~\cite{RennerThesis,MullerQuadeRenner} 
of QKD. Second, we apply large deviation
theory, specifically a multiplicative form of the Chernoff bound~\cite{cher}, to perform the parameter estimation step.
The latter is crucial to demonstrate that a long-distance 
implementation of 
mdiQKD  
(e.g., $150$ km of optical
fiber with $0.2$ dB/km) is feasible
within a reasonable time-frame. To obtain high secret key rates in this scenario, 
it is common to use decoy state techniques~\cite{decoy1,decoy2,decoy3},
both for standard QKD protocols and mdiQKD.
Here a key challenge is to estimate
the transmittance
and the quantum bit error rate (QBER) of the
single-photon component of the signal at the presence of
high losses (e.g., $30$ dB). We show that such an estimation
problem can be solved 
using the Chernoff bound, as it provides good bounds for the parameters above
even in the high-loss regime. We highlight that our results 
can be applied to other QKD protocols
(e.g., the standard decoy state BB84 protocol~\cite{decoy1,decoy2,decoy3}) as well as to general
experiments in quantum information. 

\section{Security Definition}

Prior to stating the protocol, let us quickly review the security framework~\cite{RennerThesis,MullerQuadeRenner} 
 that we are considering here. 
A general QKD protocol (executed by Alice and Bob) generates either a pair of bit strings $S_{\rm A}$ and $S_{\rm B}$, 
or a symbol $\perp$ to indicate the abort of the protocol. In general, the string of Alice, $S_{\rm A}$, can be quantum mechanically 
correlated with a quantum state that is held by the adversary. Mathematically, this situation is described by the classical-quantum state
\begin{eqnarray}
\rho_{\rm AE}=\sum_{s}\proj{s}\otimes \rho_{\rm E}^s, \nonumber
\end{eqnarray}
where $\{\ket{s}\}_s$ denotes an orthonormal basis for Alice's system, and the 
subscript ${\rm E}$ indicates the system of the adversary. 

Ideally, we say that a QKD protocol is secure if it satisfies two conditions, namely the correctness and the secrecy. 
The correctness condition is met if $S_{\rm A}=S_{\rm B}$, i.e., Alice's and Bob's bit strings are identical. 
The secrecy condition is met if $\rho_{\rm AE}=U_{\rm A} \otimes \rho_{\rm E}$, where $U_{\rm A}=\sum_s\frac{1}{|{\mathcal S}|}\proj{s}$ 
is the uniform mixture of all possible values of the bit string $S_{\rm A}$. That is, the system of the adversary is completely decoupled from 
that of Alice. 

Owing to the presence of errors, however, these two conditions can never be perfectly met. 
For example, in the finite-key regime it is impossible to guarantee $S_{\rm A}=S_{\rm B}$ with certainty. 
In practice, this implies that we need to allow for some minuscule errors. That is, we say that a QKD scheme is $\epscor$-correct 
if $\Pr[S_{\rm A} \not= S_{\rm B}] \leq \epscor$, i.e., 
the probability that Alice's and Bob's bit strings are not identical is not greater than $\epscor$.
Similarly, we say that a protocol is $\epssec$-secret if
\begin{eqnarray}
\frac{1}{2}\|\rho_{\rm AE}-U_{\rm A}\otimes\rho_{\rm E}\|_1 \leq \epssec, \nonumber
\end{eqnarray}
where $\|\cdot\|_1$ denotes the trace norm. 
That is, the state $\rho_{\rm AE}$ is $\epssec$-close to the ideal situation described by $U_{\rm A} \otimes \rho_{\rm E}$.
Thereby a QKD protocol is said to be $\epsilon$-secure if it is both $\epscor$-correct and $\epssec$-secret, 
with $\epscor+\epssec\leq\epsilon$. 

With this security definition we are able to guarantee that the security of the protocol holds even when combined with other protocols, i.e., 
the protocol is secure in the so-called universally composable framework~\cite{RennerThesis,MullerQuadeRenner}.

\section{Protocol definition}\label{sec_prot}

\begin{figure}[h]
\begin{center}
 \includegraphics[scale=0.48]{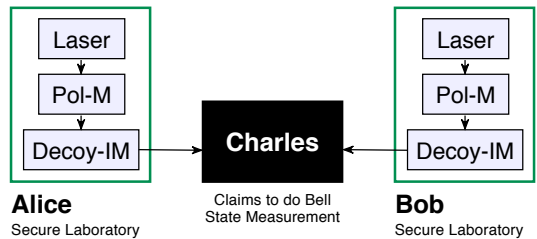}
 \end{center}
 \caption{A schematic diagram of mdiQKD.
Alice and Bob prepare quantum signals in different BB84 polarisation states~\cite{bb84}
with a polarisation modulator (Pol-M). Also, they use  
an intensity modulator (Decoy-IM) to 
generate decoy states. 
The signals are sent to an untrusted relay Charles, who is supposed to perform 
a Bell state measurement that projects the incoming signals into a Bell state. 
See the main text for details. 
\label{scheme}}
\end{figure}

The setup is illustrated in Fig.~\ref{scheme}.
Alice and Bob use a laser source to generate quantum signals that are diagonal in the Fock basis. 
Instances of such sources include attenuated laser diodes emitting phase-randomised WCPs, triggered spontaneous parametric down-conversion sources, 
and practical single-photon sources. 
Each pulse is prepared in a different BB84 state~\cite{bb84}, which is
selected, for example, uniformly at random from two mutually unbiased bases, 
denoted as $\rm Z$ and $\rm X$. 
The signals are then sent to an untrusted relay Charles, who is supposed to perform 
a Bell state measurement that projects them into a Bell state. 
Also, Alice and Bob apply decoy state techniques~\cite{decoy1,decoy2,decoy3}
to estimate the gain (i.e., the probability that the relay outputs a successful result) 
and the QBER for various input photon-numbers.

Next, Charles announces whether or not his measurements are successful, including the Bell states obtained. 
Alice and Bob keep the data that correspond to these instances and discard the rest. Also, they 
post-select the events where they employ the same basis. Finally, either Alice or Bob 
flips part of her/his bits to correctly correlate them with those of the other. See
Table~\ref{table_protocol} for a detailed description of the different steps of the protocol.
\begin{table}
\begin{description}
\item[1. State Preparation]
Alice and Bob repeat the first four steps of the protocol 
for $i=1,\ldots,N$ till the conditions in the 
Sifting step are met. 	For each $i$, Alice chooses an intensity $a\in\{a_{\rm s}, a_{\rm d_1},\ldots,a_{{\rm d}_n}\}$, 
a basis $\alpha\in\{{\rm Z, X}\}$, and a random bit $r\in\{0,1\}$ with probability
$p_{a,\alpha}/2$. Here $a_{\rm s}$ ($a_{{\rm d}_j}$) is the intensity of the signal (decoy) states.
Next, she generates a 
quantum signal (e.g., a phase-randomised WCP) 
of intensity $a$ prepared in the basis state of $\alpha$ given 
by $r$. Likewise, Bob does the same.

\item[2. Distribution] Alice and Bob send their states to Charles via the quantum channel.

\item[3. Measurement]
If Charles is honest, he measures the signals received with a Bell state measurement. 
In any case, he informs Alice and Bob (via a public channel) of whether or not his measurement was successful. If successful,
he reveals the Bell state obtained. 

\item[4. Sifting]
If Charles reports a successful result, Alice and Bob broadcast (via an authenticated channel) 
their intensity and basis settings.
For each Bell state $k$, we define two groups of sets:
${\mathcal Z}^{a,b}_k$ and ${\mathcal X}^{a,b}_k$. The first (second) one identifies 
signals
 where Charles declared the Bell state $k$ and Alice and Bob selected the
intensities $a$ and $b$ and the basis $\rm Z$
($\rm X$). 
The protocol repeats these steps until 
$|{\mathcal Z}^{a,b}_k|\geq N^{a,b}_k$ and $|{\mathcal X}^{a,b}_k|\geq M^{a,b}_k$
$\forall a, b, k$.  
Next, say Bob flips part of his bits to correctly correlate them with those of Alice
(see Table~\ref{table1}).
Afterwards, they execute the last steps of the protocol for each $k$. 

\item[5. Parameter Estimation]
Alice and Bob use $n_k$ random
bits from
${\mathcal Z}^{a_{\rm s},b_{\rm s}}_k$ 
to form the code bit strings
$Z_k$ and $Z'_k$, respectively. The 
remaining 
$R_k$ bits from ${\mathcal Z}^{a_{\rm s},b_{\rm s}}_k$ are used to compute
the error rate
$E^{a_{\rm s},b_{\rm s}}_k=\frac{1}{R_k}\sum_l r_l\oplus r'_l$,
where $r'_l$ are Bob's bits.
If $E^{a_{\rm s},b_{\rm s}}_k>E_{\rm tol}$, Alice and Bob assign an empty string to
$S_k$ and abort steps $6$ and $7$ for this
$k$.
The protocol only aborts if $E^{a_{\rm s},b_{\rm s}}_k>E_{\rm tol}$ $\forall k$.
If $E^{a_{\rm s},b_{\rm s}}_k\leq{}E_{\rm tol}$, Alice and Bob 
use ${\mathcal Z}^{a,b}_k$ and ${\mathcal X}^{a,b}_k$ to
estimate $n_{k,0}$, $n_{k,1}$ and $e_{k,1}$. The parameter $n_{k,0}$ ($n_{k,1}$)
is a lower bound for the number of bits 
in $Z_k$ where Alice (Alice and Bob) sent a vacuum (single-photon) state. 
$e_{k,1}$ is
an upper bound for the single-photon phase error rate. 
If $e_{k,1}>e_{{\rm tol}}$, 
an empty string is assigned to $S_k$ and
steps $6$ and $7$ are aborted for this $k$, and
the protocol only aborts if $e_{k,1}>e_{{\rm tol}}$ $\forall k$.

\item[6. Error Correction] For those $k$ that passed the parameters estimation step,
Bob obtains an estimate ${\hat Z}_k$ 
of $Z_k$ using 
an information reconciliation scheme. For this, Alice sends him
${\rm leak}_{{\rm EC},k}$ bits of error correction data.
Next, Alice computes a hash of $Z_k$ of length $\lceil\log_2(4/\epsilon_{\rm cor})\rceil$ using 
a random universal$_2$ 
hash function, which she sends to Bob together with the hash~\cite{RennerThesis}. 
If ${\rm hash}({\hat Z}_k)\neq {\rm hash}(Z_k)$, Alice and Bob assign an empty string to $S_k$ and abort step $7$ for this $k$. 
The protocol only aborts if 
${\rm hash}({\hat Z}_k)\neq {\rm hash}(Z_k)$
$\forall k$. 

\item[7. Privacy Amplification] If $k$ passed the error correction step,
Alice and Bob apply a random universal$_2$ hash function
to $Z_k$ and ${\hat Z}_k$ to extract two shorter strings of length $\ell_k$~\cite{RennerThesis}.
Alice obtains $S_k$ and Bob 
${\hat S}_k$. 
The concatenation of 
$S_k$ (${\hat S}_k$) form the secret key 
$S_{\rm A}$ ($S_{\rm B}$).

\end{description}
\caption{Protocol Definition.}
\label{table_protocol}
\end{table}
\begin{table}[htb]
  \centering
  \begin{tabular}{ccccc}
  \quad & \multicolumn{4}{c}{Bell state reported by Charles}\\
    \hline\hline
  Alice \& Bob & $\quad \ket{\psi^-}$ \quad & \quad $\ket{\psi^+}$ \quad & \quad $\ket{\phi^-}$ \quad & \ \quad $\ket{\phi^+}$ \ \quad \\
  \hline
  {\rm {\rm Z} basis} & \quad {\rm Bit flip}  \quad & \quad {\rm Bit flip} \quad  & \quad - \quad  & \ \quad\quad - \ \quad\quad  \\
  {\rm {\rm X} basis} & \quad {\rm Bit flip} \quad & \quad - \quad & \quad {\rm Bit flip} \quad &\ \quad\quad - \ \quad\quad \\
  \hline\hline
\end{tabular}
\caption{Post-processing of data in the sifting step. To guarantee that their bit strings are correctly correlated,
say Bob applies a bit flip to part of his data, depending on the Bell state reported by Charles 
and the basis setting selected.  
  }\label{table1}
\end{table} 

Since Charles' measurement is basically used to post-select entanglement 
between Alice and Bob, the security of 
mdiQKD can be proven by using the idea of time reversal. Indeed, mdiQKD builds on the earlier proposals of time-reversed EPR protocols 
by Biham et al.~\cite{biham} and Inamori~\cite{inamori}, and combine them with the decoy state technique. 
The end result is the best of both worlds---high performance and high security. We note on passing that the idea of 
time reversal has also been previously used in other quantum information protocols including one-way quantum computation.

\section{Security analysis}\label{sec_proof}

We now present one main result of our paper. It states that the protocol introduced above is both 
$\epscor$-correct and $\epssec$-secret, given that the length $\ell$ of the secret key 
$S_{\rm A}$
is selected appropriately for a given set of observed values. See 
Table~\ref{table_protocol} 
for the definition of the different parameters that we consider in this section.

The correctness of the protocol is guaranteed by its error correction step, where, 
for each possible Bell state $k$, 
Alice sends
a hash of $Z_k$ to Bob, who compares it with the hash of ${\hat Z}_k$. If both 
hash values are equal, the protocol gives $S_k={\hat S}_k$ except with error
probability $\epscor/4$. If ${\rm hash}({\hat Z}_k)\neq {\rm hash}(Z_k)$, it outputs the empty string
(i.e., the protocol is trivially correct). Moreover, if the protocol aborts, the result is 
$\perp$, i.e.,
it is also correct. This guarantees that $S_{\rm A}=S_{\rm B}$ except with 
error probability less or equal than $\epscor$. 
Alternatively to this method, Alice and Bob may also guarantee the
correctness of the protocol by exploiting 
properties of the error correcting code employed \cite{lut}. 

If the length $\ell_k$ of each bit string $S_k$, which forms the secret key $S_{\rm A}$, satisfies
\begin{equation}\label{klength}
\ell_k \leq n_{k,0}+n_{k,1}\left[1-h\left(e_{k,1}\right)\right]-{\rm leak}_{{\rm EC},k}-\log_2\frac{8}{\epsilon_{\rm cor}}
-2\log_2\frac{2}{\eps'_k\hat{\eps}_k}-2\log_2\frac{1}{2\eps_{k,\tn{PA}}}, 
\end{equation}
the protocol is $\epsilon_{\rm sec}$-secret, with 
$\epsilon_{\rm sec}=\sum_k \epsilon_{k,{\rm sec}}$ and $\epsilon_{k,{\rm sec}}=
2(\eps'_k+2\eps_{k,e}+\hat{\eps}_k)+\eps_{k,\tn{b}}+\eps_{k,0} 
+\eps_{k,1}+\eps_{k,\tn{PA}}$. In Eq.~(\ref{klength}), $h(x)=-x\log_2{(x)}-(1-x)\log_2{(1-x)}$ is the binary Shannon entropy, and
the parameters $\eps_{k,0} $, $\eps_{k,1}$, and $\eps_{k,e}$ quantify, respectively,
the probability that the estimation of the terms $n_{k,0}$,  $n_{k,1}$ and $e_{k,1}$ is 
incorrect. A sketch of the proof of Eq.~(\ref{klength}) can be found in Appendix~\ref{secre}.
There it is also explained the meaning of all the epsilons contained in the term 
$\epsilon_{k,{\rm sec}}$, which we omit here for simplicity.
In the asymptotic limit of very large data blocks, 
the terms reducing the length of $S_{\rm A}$
due to statistical fluctuations may be neglected, and thus $\ell$
  satisfies $\ell\leq\sum_k\max{\{n_{k,0}+n_{k,1}\left[1-h\left(e_{k,1}\right)\right]-{\rm leak}_{{\rm EC},k},0\}}$, 
as previously obtained in \cite{mdiQKD}. That is, 
$n_{k,0}$ and $n_{k,1}$ provide a positive contribution 
to the secret key rate, while $n_{k,1}h\left(e_{k,1}\right)$ and 
${\rm leak}_{{\rm EC},k}$ reduce it. 
The term $n_{k,1}h\left(e_{k,1}\right)$ corresponds to the information 
removed from $Z_k$ in the privacy amplification step of the protocol, while ${\rm leak}_{{\rm EC},k}$ is the 
information revealed by Alice in the error correction step. 

The second main contribution of this work is an estimation method to obtain the 
relevant parameters $n_{k,0}$,  $n_{k,1}$ and $e_{k,1}$ needed to evaluate the key rate formula above, 
when Alice and Bob send Charles a finite number $N$ of signals and use a finite number of decoy states. 
We solve this problem using techniques
in large deviation theory. 
More specifically, we employ
the Chernoff bound~\cite{cher}. 
It is important to note that standard techniques such as Azuma's
inequality~\cite{azuma} do not give very good bounds here. This is because this result does not consider the properties of the 
a priori distribution. Therefore, it is far from optimal for situations such as high loss or a highly bias coin flip, 
which are relevant in long-distance QKD. In contrast,
the Chernoff bound takes advantage of the property of
the distribution and provides good bounds
even in a high-loss regime. 

More precisely, 
we show that the estimation of $n_{k,0}$,
$n_{k,1}$ and $e_{k,1}$
can be 
formulated as
a linear program, 
which
 can be solved efficiently 
in polynomial time and gives the exact optimum even for large dimensions \cite{lpr}. 
Importantly, this general method is valid for any finite number of decoy states used by Alice and Bob, and
for any photon-number distribution of their signals. Also, for the typical 
scenario where 
Alice and Bob send phase-randomised WCPs together with two decoy states each, we
solve analytically the linear program, and obtain analytical
expressions for the parameters above, which can be used directly in 
an experiment. A sketch of the estimation method is given in Appendix~\ref{sketch}. For a detailed
analysis of both estimation techniques we refer to the Appendices~\ref{ap_anal} and~\ref{ap_numer}. 

\section{Discussion}

\begin{figure}[t]
\centerline{\includegraphics[width=0.24\textwidth]{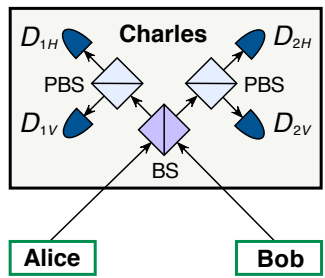}}
\vspace{-0.2cm}
\caption{A schematic diagram of Charles' measurement device.
The signals from Alice and Bob interfere at a $50:50$ beam splitter (BS),
which has on each end a polarising beam splitter (PBS) that projects
the incoming photons into either horizontal ($H$) or
vertical ($V$) polarisation states. A click in the single-photon
detectors $D_{1H}$ and 
$D_{2V}$, or in $D_{1V}$ and $D_{2H}$, indicates a projection
into the Bell state $\ket{\psi^{-}}=1/\sqrt{2}(\ket{HV}-\ket{VH})$, 
while a click in $D_{1H}$ and $D_{1V}$, or in 
$D_{2H}$ and $D_{2V}$, implies a projection into the Bell state 
$\ket{\psi^{+}}=1/\sqrt{2}(\ket{HV}+\ket{VH})$.
}\label{fig_relay}
\end{figure}

In this section we analyse the behaviour of the secret key rate 
provided in Eq.~(\ref{klength}).
In our simulation, 
we consider that Alice and Bob encode their bits in the 
polarisation degrees of freedom of phase-randomised WCPs. 
Also, we assume 
that Charles uses the linear optics quantum relay 
illustrated in Fig.~\ref{fig_relay},
which is able to 
identify 
two of the four Bell states. 
With this setup, a successful Bell state measurement 
corresponds to the observation of precisely two detectors 
(associated to orthogonal polarisations) being triggered.
Note, however, that the results presented in this paper can be 
applied to other 
types of coding schemes like, for instance, 
phase or time-bin coding~\cite{QKDreview1,QKDreview2}, 
and to
any quantum operation that Charles may perform, 
as they solely depend on the measurement results that he announces. 

We use experimental parameters from~\cite{param}.
But, whereas~\cite{param} considers a free-space channel,
we assume
a fiber-based channel with a loss of $0.2$ dB/km. 
The detection efficiency of the 
relay (i.e., the transmittance of its optical components together with the efficiency of its 
detectors)
is $14.5\%$, and the 
background count rate is $6.02\times{}10^{-6}$. 
Moreover, we use a rather generic channel model 
that includes an intrinsic error rate which
simulates the misalignment and instability of the optical system.
This is done by placing a unitary rotation in both input arms of the $50:50$ beam splitter, 
and another unitary rotation in one of its output arms \cite{feihu}. 
In addition, 
we 
fix the security bound to
$\epsilon=10^{-10}$.
\begin{figure}[t]
\centerline{\includegraphics[width=0.44\textwidth]{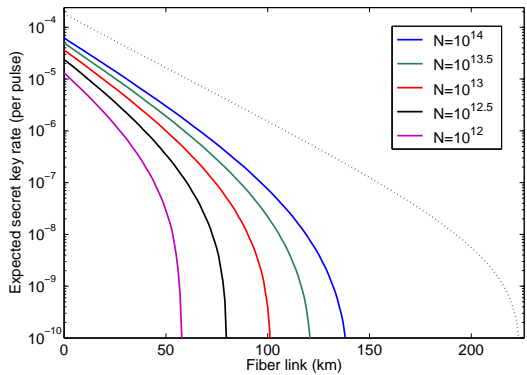}}
\vspace{-0.2cm}
\caption{Expected key rate as function of the distance. Secret key rate $\ell/N$ in logarithmic scale for the 
protocol introduced in section~\ref{sec_prot} with phase-randomised WCPs as a
function of the distance. The solid lines correspond to different values for the total number of signals $N$
sent by Alice and Bob. The overall misalignment 
in the channel is $1.5\%$, 
and the security bound $\epsilon=10^{-10}$. 
For simulation purposes we consider the following experimental 
parameters~\cite{param}: the loss coefficient of the channel is 
$0.2$ dB/km, the detection efficiency of the relay is $14.5\%$, 
and the 
background count rate is $6.02\times{}10^{-6}$. 
Our results show clearly that even with a realistic finite size of data, 
say $N=10^{12}$ to $10^{14}$, it is possible to 
achieve secure mdiQKD at long distances.
In comparison, the dotted line represents a lower bound on the 
secret key rate for the asymptotic case where Alice and Bob send 
Charles
infinite signals and use an infinite number of decoy settings. 
}\label{fig1}
\end{figure}
\begin{figure}[t]
\centerline{\includegraphics[width=0.44\textwidth]{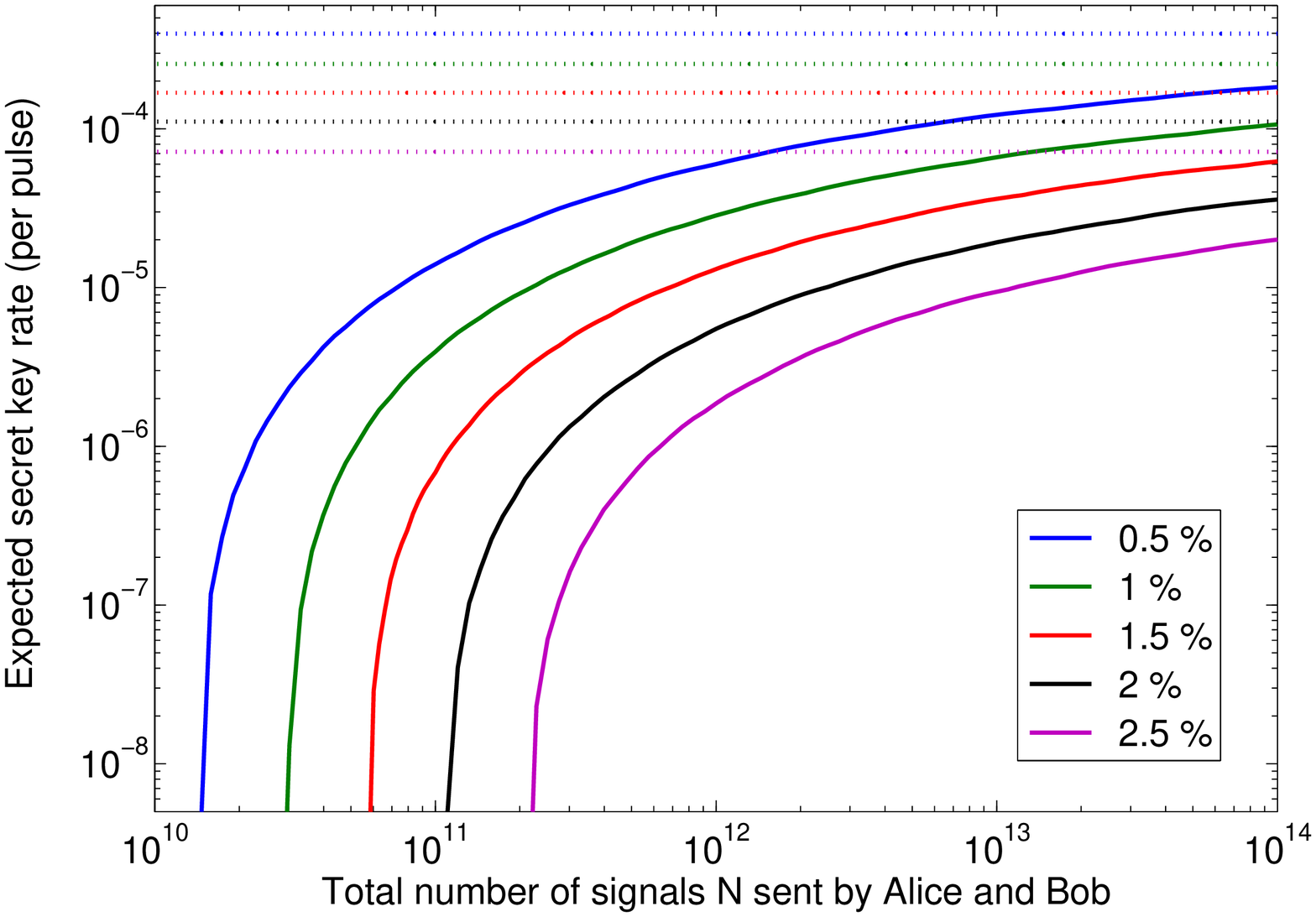}}
\vspace{-0.2cm}
\caption{Expected key rate as function of the block size. The plot shows the secret key rate $\ell/N$ in logarithmic scale
as a function of the total number of signals $N$
sent by Alice and Bob in the limit of zero distance. The security bound $\epsilon=10^{-10}$. 
The solid lines correspond to different values for the intrinsic error 
rate due to the misalignment and instability of the optical system.
The horizontal dotted lines show the asymptotic rates.
The experimental parameters are the ones described in the 
caption of Fig.~\ref{fig1}.
Our results show that, even for a finite size of signals sent 
by Alice and Bob, mdiQKD is robust to intrinsic errors due to basis 
misalignment and instability of the optical system.}\label{fig2}
\end{figure}
 
The results are shown in Figs.~\ref{fig1}-\ref{fig2} for the situation where Alice and Bob use two decoy states each. 
In this scenario, 
we obtain 
the 
parameters $n_{k,0}$, $n_{k,1}$ and $e_{k,1}$ using the analytical estimation procedure introduced above 
(see Appendix~\ref{ap_anal} for more details).
The first figure illustrates the secret key rate (per pulse) $\ell/N$ as a function of the distance
between Alice and Bob 
for different values of the total number of signals $N$ sent. 
We fix $\epsilon_{\rm cor}=10^{-15}$; this corresponds to a realistic
hash tag size in practice \cite{RennerThesis}. 
Also, we fix
the intensity of the weakest 
decoy states to $a_{\rm d_2}=b_{\rm d_2}=5\times10^{-4}$, since, in practice, is 
difficult to generate a vacuum state due to imperfect extinction. This value for $a_{\rm d_2}$ and $b_{\rm d_2}$ can be easily achieved with 
a standard intensity modulator. 
Moreover, for simplicity, we assume an 
error correction leakage that is a fixed fraction of the sifted key length $n_k$, i.e.,
${\rm leak}_{{\rm EC},k}=n_k\zeta{}h(E_k^{a_{\rm s},b_{\rm s}})$, 
with $\zeta=1.16$ and 
where 
$h(\cdot)$
is again the binary Shannon entropy \cite{finite1}.
In a realistic scenario, however, the value of $\zeta$ typically depends on the 
value of $n_k$, and when $n_k< 10^5$ the parameter 
$\zeta$ may be bigger than $1.16$. 
For a given distance, we optimise numerically $\ell/N$ over all the 
free parameters of the protocol. This includes the intensities 
$a_{\rm s}, a_{\rm d_1}, b_{\rm s}$ and $b_{\rm d_1}$,
the probability 
distributions $p_{a,\alpha}$ and $p_{b,\beta}$
in the state preparation step, the parameters 
$N_k^{a,b}$ and $M_k^{a,b}$ in the sifting step, the term
$n_k$ in the parameter estimation step,
and the different epsilons contained in $\epsilon_{\rm sec}$. 
Our simulation result shows clearly that mdiQKD is feasible
with current technology and does not
require high efficiency detectors for its implementation.
If Alice and Bob use laser diodes operating at $1$ GHz repetition rate,
and each of them sends $N=10^{13}$ signals,
we find, for instance, that they can distribute a 
$1$ Mb secret key over a $75$ km fiber link in less than $3$ hours. 
This scenario corresponds to the
red line shown in Fig.~\ref{fig1}. 
Notice that, at telecom wavelengths, standard InGaAs detectors have
modest detection efficiency of about $15\%$. Since mdiQKD requires
two-fold coincidence rather than single-detection events, as is the case in the
standard decoy state protocol, the key rate of mdiQKD is
lower than that of the standard decoy state scheme. However,
with high-efficiency detectors such as silicon detectors~\cite{det1}
in $800$nm or high-efficiency SSPDs~\cite{det2}, the key rate of
mdiQKD can be made comparable to that of the standard decoy state protocol.

The second figure illustrates $\ell/N$ as a function of $N$ for different 
values of the misalignment in the limit of zero distance. 
For comparison, this figure also includes the asymptotic secret key rate 
when Alice and Bob send an infinite number of signals and
use an infinite number of decoy states \cite{mdiQKD}. 
Our results show that significant secret key rates are already possible with
$10^{11}$ signals, given that the error rate is not too large. 

\section{Conclusion}

We have proved the security of 
mdiQKD in the finite-key regime against general attacks. 
This is the only known
fully practical QKD protocol that offers an avenue to bridge the 
gap between theory and practice in QKD implementations. 
Importantly, our results clearly demonstrate that even with practical 
signals [e.g., phase-randomised weak coherent pulses (WCPs)] and 
a finite size of data (say $10^{12}$ to $10^{14}$ signals) it is possible to 
perform secure mdiQKD over long distances (up to about $150$ km). 

To achieve high secret key 
rates in such high-loss regime, it is typical 
both for standard QKD schemes and
mdiQKD
to use decoy state techniques. A main challenge in this scenario is to obtain tight bounds 
for the gain and quantum bit error rate (QBER) of the single-photon components
sent by Alice and Bob. We have shown that this estimation problem can be successfully 
solved using techniques in large deviation theory, more precisely, 
the Chernoff bound. This result takes advantage of the property of the 
distribution, and thus provides good bounds for the relevant 
parameters even in the presence of high losses, as is the case in QKD realisations. 

Using the Chernoff bound we have rewritten the problem of
estimating the gain and QBER of the single-photon signals 
as a linear program, 
which
 can be solved efficiently 
in polynomial time. This general method is valid for any finite number of decoy states, and
for any photon-number distribution of the signals. It can be used, 
for instance, with
laser diodes emitting phase-randomised WCPs, triggered spontaneous parametric 
down-conversion sources, 
and practical single-photon sources. 
Also, for the common
scenario where 
Alice and Bob send phase-randomised WCPs together with two decoy states each, we
have obtained tight analytical bounds for the quantities above. These results apply 
to different types of coding schemes like, for example, 
polarisation, phase or time-bin coding.

\section{Acknowledgements}

We thank Xiongfeng Ma and Johan L\"ofberg for valuable comments
and stimulating discussions, and 
Lina M. Eriksson
for comments on the writing and presentation of the paper. 
F. Xu. thanks the Paul Biringer Graduate Scholarship for financial support.
We
acknowledge support from the European Regional Development Fund (ERDF),
the Galician Regional Government (projects CN2012/279 and CN 2012/260,
``Consolidation of Research Units: AtlantTIC"), 
NSERC, the CRC program, 
the
National Centre of Competence in Research QSIT, 
the Swiss NanoTera project QCRYPT, the FP7 Marie-Curie IAAP
QCERT project, and CHIST-ERA project DIQIP.
K.T. acknowledges support from the project ``Secure photonic network technology''
as part of ``The project UQCC'' by the National
Institute of Information and Communications Technology
(NICT) of Japan, and from the Japan Society for
the Promotion of Science (JSPS) through its Funding
Program for World-Leading Innovative R\&D on Science
and Technology (FIRST Program).

\appendix

\section{Secrecy}\label{secre}

Here, we briefly discuss on the secrecy of the protocol described in 
Table~\ref{table_protocol}.
To begin with, note that Alice and Bob obtain
the error rate $E^{a_{\rm s},b_{\rm s}}_k$ using a random sample of 
${\mathcal Z}_k^{a_{\rm s},b_{\rm s}}$ of 
size $R_k$. This means that 
when $E^{a_{\rm s},b_{\rm s}}_k$ satisfies the tolerated value $E_{\tn{tol}}$,
the error rate between the strings 
$Z_k$ and $Z'_k$, which we denote as $\xi^{a_{\rm s},b_{\rm s}}_k$,
satisfies the following inequality written as conditional probability \cite{serf}
\be \label{PE1}
\Pr\left[ \xi^{a_{\rm s},b_{\rm s}}_k \geq  E^{a_{\rm s},b_{\rm s}}_k + \chi(n_k,R_k,{\bar \eps}_k)| \Omega_\tn{pass}\right] \leq {\bar \eps}_k^2,
\ee
where 
$\chi(x,y,z)=\sqrt{(y+x)(y+1)/(xy^2)\ln{}z^{-1}}$.
Here, the parameter $\Omega_{\tn{pass}}$ represents the event that all the tests performed during the realisation of the 
protocol satisfy the tolerated values.

Let $E'_k$ denote the adversary's information about $Z_k$ up to the error correction step in Table~\ref{table_protocol}. 
By using a privacy amplification scheme based on two-universal hashing~\cite{RennerThesis}
we can generate an $\epsilon_k$-secret string $S_k$ of length $\ell_k$, where $\eps_k > 0$, and 
\be \label{Lemma:PA}
\epsilon_k \leq 8\eps_k + 2^{-\frac{1}{2}\left(\HminOp^{4\eps_k}\left(Z_k|E'_k\right)-\ell_k\right)-1}. 
\ee
The function $\HminOp^{4\eps_k}\left(Z_k|E'_k\right)$ denotes the smooth min-entropy~\cite{RennerThesis,minentr}.
It quantifies the average probability that the adversary guesses $Z_k$ correctly using the optimal 
strategy with access to $E'_k$.

The term $E'_k$ can be decomposed as  
$E'_k=C_kE_k$, where $C_k$ is the 
information
revealed by Alice and Bob during the 
error correction step, and $E_k$ is the adversary's information prior to that step. 
Using a chain-rule for smooth entropies~\cite{RennerThesis}, we obtain
\be \label{CR1}
\HminOp^{4\eps_k}\left(Z_k|E'_k\right) \geq \HminOp^{4\eps_k}\left(Z_k|E_k\right)-|C_k|,
\ee 
with $|C_k|\leq{\rm leak}_{{\rm EC},k}+\log_2(8/\epsilon_{\rm cor})$.

The bits of $Z_k$ can be distributed among three different strings: 
$Z_k^0$, $Z_k^1$ and $Z_k^{\rm rest}$. 
The first contains bits where Alice sent a vacuum state, the second 
where both Alice and Bob sent a single-photon state, and $Z_k^{\rm rest}$ includes the rest of bits. 
Using a result from~\cite{Tomamichel}, we have that
\begin{equation} \label{CR2}
\HminOp^{4\eps_k}\left(Z_k|E_k\right)\geq
\HminOp^{\eps'_k+2\eps''_k+(\hat{\eps}_k+2\hat{\eps}'_k+\hat{\eps}''_k)}
\left(Z_k^0Z_k^1Z_k^{\rm rest}|E_k\right) 
\geq
n_{k,0}+
\HminOp^{\eps''_k}\left(Z_k^1|Z_k^0Z_k^{\rm rest}E_k\right)
-2\log_2\frac{2}{\eps'_k\hat{\eps}_k}, 
\end{equation}
where $4\eps_k=\eps'_k+2\eps''_k+(\hat{\eps}_k+2\hat{\eps}'_k+\hat{\eps}''_k)$.
Here, we have used the fact that
$\HminOp^{\hat{\eps}'_k}\left(Z_k^{\rm rest}|Z_k^0E_k\right)\geq 0$, and 
$\HminOp^{\hat{\eps}''_k}\left(Z_k^0|E_k\right) \geq \HminOp^{0}\left(Z_k^0|E_k\right)
=\HminOp \left(Z_k^0\right)=n_{k,0}$. The latter arises because vacuum states contain no information 
about their bit values, which are uniformly distributed.

The next step is to obtain a lower bound for the term $\HminOp^{\eps''_k}\left(Z_k^1|Z_k^0Z_k^{\rm rest}E_k\right)$.
Taking that Alice and Bob do the state preparation scheme perfectly in the $\rm Z$ and $\rm X$ bases
(i.e., they prepare perfect BB84 states),
 we can re-write this quantity in terms of the smooth max-entropy between them, which is directly bounded by 
 the strength of their correlations \cite{finite1}. More precisely, the entropic uncertainty relation gives us
\begin{equation}  \label{UCR}
\HminOp^{\eps''_k}\left(Z_k^1|Z_k^0Z_k^{\rm rest}E_k\right)  
\geq n_{k,1}-\HmaxOp^{\eps''_k}\left(X_k^1|X'_k{}^{1}\right) 
\geq n_{k,1}-n_{k,1}h\left(e_{k,1}\right).
\end{equation}

Combining Eqs.~(\ref{Lemma:PA},\ref{CR1},\ref{CR2},\ref{UCR}), we find that a 
secret key of length $\ell_k$ given by Eq.~(\ref{klength})
gives an error of 
$\epsilon_k \leq  2(\eps'_k+2\eps''_k+\hat{\eps}_k+2\hat{\eps}'_k+\hat{\eps}''_k) + \eps_{k,\tn{PA}}$. 
Finally, after composing the errors 
related with the estimation of $n_{k,0}$, $ n_{k,1}$ and $e_{k,1}$, 
selecting $\hat{\eps}'_k$ and $\hat{\eps}''_k$ equal to zero,
and also removing the conditioning on $\Omega_{\tn{pass}}$, we obtain a security parameter $\epsilon_{k,{\rm sec}}$ given by
\begin{equation}
\epsilon_{k,{\rm sec}}=
2(\eps'_k+2\eps_{k,e}+\hat{\eps}_k)+\eps_{k,\tn{b}}+\eps_{k,0}+\eps_{k,1} 
+\eps_{k,\tn{PA}},
\end{equation}
where $\eps_{k,\tn{b}}={\bar \eps}_k\sqrt{\Pr[\Omega_\tn{pass}]}$, and $\eps_{k,0}$, $\eps_{k,1}$ 
and $\eps_{k,e}$ denote, respectively,
the error probability in the estimation of $ n_{k,0}$, $ n_{k,1}$ and $e_{k,1}$.

\section{Sketch of the parameter estimation method}\label{sketch} 

To simplify the discussion, let us consider the estimation of the parameter $n_{k,0}$.
The method to obtain $n_{k,1}$ and $e_{k,1}$ follows similar arguments. 
The procedure can be divided into two steps. First, 
we calculate
a lower bound for the number of indexes in ${\mathcal Z}^{a_{\rm s},b_{\rm s}}_k$ 
where Alice sent a vacuum state. This quantity is denoted 
as
$m_{k,0}$.
Second, we compute $n_{k,0}$ from $m_{k,0}$ using 
the Serfling inequality for random sampling without replacement \cite{serf}. 

In the first step we use a multiplicative form of the Chernoff bound~\cite{cher}
for independent random variables, 
which does not require the prior knowledge on the population mean.
More precisely, we use the following Claim. 

\noindent {\bf Claim 1}: Let 
$X_1, X_2, \ldots, X_n$, be a set of independent Bernoulli random variables that 
satisfy ${\rm Pr}(X_i=1)=p_i$, and let $X=\sum_{i=1}^n X_i$ and
$\mu=E[X]=\sum_{i=1}^n p_i$, 
where $E[\cdot]$ denotes the mean value. 
Let $x$ be the observed outcome of $X$ for a given trial (i.e., $x\in{\mathbb N}^+$)
and $\mu_{\rm L}=x-\sqrt{n/2\ln{(1/\epsilon)}}$ for 
certain $\epsilon>0$. When 
$(2\varepsilon^{-1})^{1/\mu_{\rm L}}\leq\exp{\left[3/(4\sqrt{2})\right]^2}$ and
$({\hat \varepsilon}^{-1})^{1/\mu_{\rm L}}<\exp{(1/3)}$ 
for certain $\varepsilon, {\hat \varepsilon}> 0$,
we have that $x$ satisfies
\begin{equation}\label{cher_met}
x=\mu+\delta,
\end{equation}
except with error probability $\gamma=\epsilon+\varepsilon+{\hat \varepsilon}$, where the 
parameter $\delta\in[-\Delta,{\hat \Delta}]$, with 
$\Delta=g(x, \varepsilon^4/16)$,
${\hat \Delta}=g(x, {\hat \varepsilon}^{3/2})$ and $g(x,y)=\sqrt{2x\ln{(y^{-1}})}$.
Here $\varepsilon$ (${\hat \varepsilon}$) denotes the 
probability that $x<\mu-\Delta$ ($x>\mu+{\hat \Delta}$).

Importantly, the bounds ($-\Delta$ and ${\hat \Delta}$) on the fluctuation parameter 
$\delta$ that appears in Eq.~(\ref{cher_met}) do not depend on the mean value $\mu$. A proof of 
Claim $1$ can be found in Appendix~\ref{chernoff}. There, we introduce as well a generalised
version of Claim $1$ for the cases where  $(2\varepsilon^{-1})^{1/\mu_{\rm L}}>\exp{\left[3/(4\sqrt{2})\right]^2}$ and/or
$({\hat \varepsilon}^{-1})^{1/\mu_{\rm L}}\geq\exp{(1/3)}$. 

In order to apply this statement and be able to obtain 
the parameter $m_{k,0}$, 
we rephrase the protocol 
described in section~\ref{sec_prot} as follows. 
For each signal, we consider that Alice (Bob) first chooses a photon-number $n$ ($m$)
and sends the signal to Charles, who declares whether his measurement is
successful or not.
After, Alice decides the intensity setting $a$, and Bob does the same. 
This virtual protocol is equivalent to the original one because the essence
of decoy state QKD is precisely that Alice and Bob could have postponed
the choice of which states are signals or
decoys after Charles' declaration of the successful events. 
This is possible because Alice's and Bob's observables 
commute with those of Charles. Note that for each specific
combination of values $n$ and $m$, the observables that 
Alice and Bob use to determine whether a state is a signal or a decoy act on entirely different 
physical systems from those of Charles. This implies that Alice and Bob are free to postpone 
their measurement and thus their choice of  
signals and decoys. Also, this result shows that 
for each combination $n$ and $m$, 
the signal and decoy states provide 
a random sample of the population of all signals containing $n$ and $m$ photons respectively. 
Therefore, one can 
apply random sampling theory in classical statistics to the quantum problem. 

Let ${\mathcal S}_{k,nm}$ denote 
the set that identifies those signals sent by Alice and Bob with 
$n$ and $m$ photons respectively, when they select 
the $\rm Z$ basis and Charles announces the Bell state $k$. 
And, let
$|{\mathcal S}_{k,nm}|=S_{k,nm}$, and
$p_{a,b|nm,{\rm Z}}$ be the conditional probability that Alice and Bob have selected 
the intensity settings $a$ and $b$, given that their signals contain, respectively, 
$n$ and $m$ photons prepared in the ${\rm Z}$ basis.
Then, if we apply the above equivalence, 
independently of each other and for each signal
Alice and Bob assign to each
element in ${\mathcal S}_{k,nm}$
the intensity setting $a, b$, 
 with 
probability $p_{a,b|nm,{\rm Z}}$. 

Let $X_{i|k,nm}^{a,b}$ be $1$ if the 
$i$th element of ${\mathcal S}_{k,nm}$ is assigned to the 
intensity setting combination $a,b$, and otherwise
$0$. And, let 
\begin{equation}
X_k^{a,b}=\sum_{n,m}\sum_{i=1}^{S_{k,nm}}X_{i|k,nm}^{a,b},
\end{equation}
with $\mu_k^{a,b}=E[X_k^{a,b}]=\sum_{n,m}p_{a,b|nm,{\rm Z}}S_{k,nm}$. 
Let $x_k^{a,b}=|{\mathcal Z}^{a,b}_k|$ denote the observed outcome of the random variable 
$X_k^{a,b}$ for a given trial.
Then, if
$(2\varepsilon_{a,b}^{-1})^{1/\mu_{k,{\rm L}}^{a,b}}\leq\exp{\left[3/(4\sqrt{2})\right]^2}$ and
$({\hat \varepsilon}_{a,b}^{-1})^{1/\mu_{k,{\rm L}}^{a,b}}<\exp{(1/3)}$, with
\begin{equation}
\mu_{k,{\rm L}}^{a,b}=|{\mathcal Z}^{a,b}_k|
-\sqrt{\sum_{a,b}\vert{\mathcal Z}_k^{a,b}\vert/2\ln{(1/\epsilon_{a,b})}},
\end{equation}
the 
Claim above implies that 
\begin{equation}\label{g5method}
|{\mathcal Z}^{a,b}_k|
=\sum_{n,m}p_{a,b|nm,{\rm Z}}S_{k,nm}+\delta_{a,b},
\end{equation}
except with error probability $\gamma_{a,b}=\epsilon_{a,b}+\varepsilon_{a,b}+{\hat \varepsilon}_{a,b}$, 
where  
$\delta_{a,b}\in[-\Delta_{a,b},{\hat \Delta}_{a,b}]$, with  
$\Delta_{a,b}=g(|{\mathcal Z}^{a,b}_k|,\varepsilon_{a,b}^4/16)$ and
${\hat \Delta}_{a,b}=g(|{\mathcal Z}^{a,b}_k|,{\hat \varepsilon}_{a,b}^{3/2})$. 

Using similar arguments, we find that the parameter $m_{k,0}$ can be written as
\begin{equation}\label{mk0_method}
m_{k,0}={}\sum_{m}
p_{a_{\rm s},b_{\rm s}|0m,{\rm Z}}S_{k,0m}-\Delta_{0},
\end{equation}
except with error probability $\varepsilon_{0}$, where 
$\Delta_{0}=g(\sum_{m}
p_{a_{\rm s},b_{\rm s}|0m,{\rm Z}}S_{k,0m},\varepsilon_{0})$. 

Now, it is easy to find a lower bound for $m_{k,0}$. 
One only needs to minimise Eq.~(\ref{mk0_method})
 given the 
linear constraints imposed by Eq.~(\ref{g5method}) for all $a, b$. 
This problem can be solved either using numerical tools as linear programming~\cite{lpr} or, 
for some particular cases, also analytical techniques. See 
the Appendices~\ref{ap_anal} and~\ref{ap_numer} for details. 

The second step of the procedure is quite direct.  
Note that Alice 
forms her bit string $Z_k$ using $n_k$ random indexes from ${\mathcal Z}_k^{a_{\rm s},b_{\rm s}}$. 
Using~\cite{serf} we obtain
\begin{equation}\label{eq_in_0method}
n_{k,0}=
\max\bigg\{
\bigg\lfloor{}n_{k}\frac{m_{k,0}}{|{\mathcal Z}^{a_{\rm s},b_{\rm s}}_k|}-n_{k}
\Lambda(|{\mathcal Z}^{a_{\rm s},b_{\rm s}}_k|,n_k,\varepsilon''_{k,0}) \bigg\rfloor,0\bigg\},
\end{equation} 
except with error probability 
\begin{equation}\label{eps_0method}
\varepsilon_{k,0}\leq \varepsilon'_{k,0}+\varepsilon''_{k,0},
\end{equation}
where $\varepsilon'_{k,0}$ corresponds to the total error probability in the estimation of 
$m_{k,0}$, and the function $\Lambda(x,y,z)$ is defined as
$\Lambda(x,y,z)=\sqrt{(x-y+1)\ln{(z^{-1})}/(2xy)}$.

\section{Analytical estimation of 
$n_{k,0}$,
$n_{k,1}$ and $e_{k,1}$}\label{ap_anal}

This Appendix contains a general method to obtain an analytical 
expression for $n_{k,0}$,
$n_{k,1}$ and $e_{k,1}$, when Alice and Bob use two decoy states each
and the photon-number distribution of their signals is Poissonian.

That is, here we assume that 
$a\in{\mathcal A}=\{a_{\rm s},a_{{\rm d}_1},a_{{\rm d}_2}\}$, with
$a_{\rm s}>a_{{\rm d}_1}>a_{{\rm d}_2}$, 
$b\in{\mathcal B}=\{b_{\rm s},b_{{\rm d}_1},b_{{\rm d}_2}\}$, with
$b_{\rm s}>b_{{\rm d}_1}>b_{{\rm d}_2}$, and the probability 
that Alice (Bob) sends an $n$-photon ($m$-photon) signal 
when she (he) selects the intensity $a$ ($b$) is given by
$p_{n|a}=e^{-a}a^n/n!$ ($p_{m|b}=e^{-b}b^m/m!$).

A similar estimation procedure has been recently introduced in~\cite{Wang:2013}.
Note, however, that~\cite{Wang:2013} considers one of the two decoy signals
a vacuum state, which is very hard to guarantee in practical QKD implementations, 
due to the finite extinction ratio of the intensity modulator~\cite{decoy:2009}.
Also, \cite{Wang:2013} analyses the asymptotic 
regime where Alice and Bob send an arbitrarily large number of signals. 
Below we introduce a general 
analytical method that overcomes both difficulties. 

We begin by introducing some notations. 
Let $S_{k,nm}$
denote the number of signals sent by 
Alice and Bob with $n$ and $m$ photons respectively, when they select 
the basis ${\rm Z}$ and 
Charles declares the 
Bell state $k$. 
As noted in Appendix~\ref{sketch}, for each combination of values $n$ and $m$, the signal and decoy states 
provide a random sample of the population of all signals containing $n$ and $m$ photons
respectively. Therefore, standard large deviation theory techniques such as the 
Chernoff bound apply~[34]. In particular, when both
\begin{eqnarray}\label{cond_sun}
\left(2\varepsilon_{a,b}^{-1}\right)^{1/\mu_{k,{\rm L}}^{a,b}}&\leq&e^{\left[3/(4\sqrt{2})\right]^2}, \nonumber \\ 
\left({\hat \varepsilon}_{a,b}^{-1}\right)^{1/\mu_{k,{\rm L}}^{a,b}}&<&e^{1/3}, 
\end{eqnarray}
with the parameter $\mu_{k,{\rm L}}^{a,b}$ given by
\begin{equation}\label{bfff}
\mu_{k,{\rm L}}^{a,b}=|{\mathcal Z}^{a,b}_k|
-\sqrt{\sum_{a,b}\vert{\mathcal Z}_k^{a,b}\vert/2\ln{(1/\epsilon_{a,b})}},
\end{equation}
we have that $|{\mathcal Z}^{a,b}_k|$ can be written as
\begin{equation}\label{g5}
|{\mathcal Z}^{a,b}_k|
=\sum_{n,m}p_{a,b|nm,{\rm Z}}S_{k,nm}+\delta_{a,b},
\end{equation}
except with error probability $\gamma_{a,b}=\epsilon_{a,b}+\varepsilon_{a,b}+{\hat \varepsilon}_{a,b}$,
where $\varepsilon_{a,b}$ refers to the failure probability of one side, whereas 
${\hat \varepsilon}_{a,b}$ refers to that of the other side. 
The total failure probability is thus the sum and is denoted by $\gamma_{a,b}$.
The parameter
$\delta_{a,b}\in[-\Delta_{a,b},{\hat \Delta}_{a,b}]$, with  
$\Delta_{a,b}=g(|{\mathcal Z}^{a,b}_k|,\varepsilon_{a,b}^4/16)$ and
${\hat \Delta}_{a,b}=g(|{\mathcal Z}^{a,b}_k|,{\hat \varepsilon}_{a,b}^{3/2})$, and
the function
$g(x,y)=\sqrt{2x\ln{(y^{-1}})}$.
A proof of Eq.~(\ref{g5}) can be found in Appendix~\ref{sketch}
(see also Appendix~\ref{chernoff}), 
where we introduce as well a generalised
version of it for the cases where Eq.~(\ref{cond_sun}) is not satisfied. 
 
\subsection{Estimation of $n_{k,0}$}\label{sub_n11b}

The procedure to obtain $n_{k,0}$ can be decomposed into two steps. First, we calculate
a lower bound for the number of indexes in ${\mathcal Z}^{a_{\rm s},b_{\rm s}}_k$ 
where Alice sent a vacuum state. This quantity is denoted 
as
$m_{k,0}$, and can be written as
\begin{equation}\label{mk0}
m_{k,0}={}\sum_{m}
p_{a_{\rm s},b_{\rm s}|0m,{\rm Z}}S_{k,0m}-\Delta_{0},
\end{equation}
except with error probability $\varepsilon_{0}$, where 
$\Delta_{0}=g(\sum_{m}
p_{a_{\rm s},b_{\rm s}|0m,{\rm Z}}S_{k,0m},\varepsilon_{0})$. 
The proof of Eq.~(\ref{mk0}) follows similar lines as the proof of Eq.~(\ref{g5})~[34].
Second, we use the Serfling inequality for random sampling without replacement~[49] to 
compute $n_{k,0}$ from $m_{k,0}$. 

Let us begin with the first step. According to Eq.~(\ref{mk0}), to compute $m_{k,0}$ 
we need to search
for a lower bound for 
\begin{equation}\label{q1a}
\sum_m p_{a_{\rm s},b_{\rm s}|0m,{\rm Z}}S_{k,0m}.
\end{equation} 
The probability $p_{a,b|nm,{\rm Z}}$ can be written as
\begin{equation}
p_{a,b|nm,{\rm Z}}=\frac{p_{n|a}p_{m|b}p_{a,b,{\rm Z}}}{\sum_{a,b}p_{n|a}p_{m|b}p_{a,b,{\rm Z}}}, 
\end{equation}
where
$p_{a,b,{\rm Z}}$
denotes the probability that Alice and Bob send signals in the $\rm Z$ basis 
with intensity $a$ and $b$ respectively. Using the fact that
$p_{n|a}=e^{-a}a^n/n!$ and $p_{m|b}=e^{-b}b^m/m!$ we obtain
\begin{equation}\label{q2a}
p_{a,b|nm,{\rm Z}}=\frac{1}{n!m!\tau_{nm}}e^{-(a+b)}a^nb^mp_{a,b,{\rm Z}},
\end{equation}
with the term $\tau_{nm}$ given by
\begin{equation}\label{taoo}
\tau_{nm}=\frac{1}{n!m!}\sum_{a,b} e^{-(a+b)}a^nb^mp_{a,b,{\rm Z}}.
\end{equation}
Hence, we have that Eq.~(\ref{q1a}) can be expressed as 
\begin{equation}
p_{a_{\rm s},b_{\rm s},{\rm Z}}e^{-(a_{\rm s}+b_{\rm s})}T_{k,0m},
\end{equation} 
with the parameter $T_{k,0m}$ given by
\begin{equation}
T_{k,0m}=\sum_m \frac{b_{\rm s}^m}{m!}{\tilde S}_{k,0m}, 
\end{equation}
where ${\tilde S}_{k,nm}=S_{k,nm}/\tau_{nm}$. In so doing, we reduce the problem of finding $m_{k,0}$ to that of 
calculating a lower bound for $T_{k,0m}$. This is what we do next. 

Our starting point is Eq.~(\ref{g5}), which now can be rewritten as
\begin{equation}\label{monday}
|{\tilde {\mathcal Z}}^{a,b}_k|
=\sum_{n,m}\frac{a^nb^m}{n!m!}{\tilde S}_{k,nm}+{\tilde \delta}_{a,b},
\end{equation}
with $|{\tilde {\mathcal Z}}^{a,b}_k|=e^{a+b}|{\mathcal Z}^{a,b}_k|/p_{a,b,{\rm Z}}$ 
and 
${\tilde \delta}_{a,b}=e^{a+b}\delta_{a,b}/p_{a,b,{\rm Z}}$.
Next, we combine the quantities given by Eq.~(\ref{monday}) in such a way that we can cancel out
the terms of the form ${\tilde S}_{k,1m}$. For this, we
define the parameter $L_{k,a_0,a_1}$ as
 \begin{equation}\label{ghj}
L_{k,a_0,a_1}=a_0|{\tilde {\mathcal Z}}^{a_1,b_{\rm s}}_k|-a_1|{\tilde {\mathcal Z}}^{a_0,b_{\rm s}}_k|
=(a_0-a_1)T_{k,0m} +\sum_{\substack{m=0\\ n\geq 2 }}\frac{a_0a_1^n-a_1a_0^n}{n!m!}b_{\rm s}^m{\tilde S}_{k,nm}
+a_0{\tilde \delta}_{a_1,b_{\rm s}}-a_1{\tilde \delta}_{a_0,b_{\rm s}},
 \end{equation}
with $a_0, a_1\in{\mathcal A}$. 
Note that when $a_0>a_1$
the second term on the r.h.s. of
 Eq.~(\ref{ghj}) is always less or equal to zero. This means, in particular,
 that $L_{k,a_0,a_1}\leq(a_0-a_1)T_{k,0m} 
 +a_0{\hat \Gamma}_{a_1,b_{\rm s}}+a_1{\Gamma}_{a_0,b_{\rm s}}$. 
 Here, for the fluctuation terms ${\tilde \delta}_{a,b}$, we have used the fact that 
 they lay in the interval $[-\Gamma_{a,b}, {\hat \Gamma}_{a,b}]$, 
with $\Gamma_{a,b}=e^{a+b}\Delta_{a,b}/p_{a,b,{\rm Z}}$ and 
${\hat \Gamma}_{a,b}=e^{a+b}{\hat \Delta}_{a,b}/p_{a,b,{\rm Z}}$, except with 
error probability $\gamma_{a,b}$.

As a result we find, therefore, that 
 \begin{equation}
T_{k,0m} \geq
 \max_{\substack{a_0,a_1\in{\mathcal A}\\ a_0>a_1}}
 \left\{
 \frac{L_{k,a_0,a_1}
 -a_0{\hat \Gamma}_{a_1,b_{\rm s}}-a_1{\Gamma}_{a_0,b_{\rm s}}}
 {a_0-a_1},0\right\},
 \end{equation}
 except with error probability $\sum_{a}\gamma_{a,b_{\rm s}}$.
 
Moving to the second step, we use the Serfling inequality~[49] to compute $n_{k,0}$ from $m_{k,0}$. 
This is so because Alice 
forms her bit string $Z_k$ using $n_k$ random indexes from ${\mathcal Z}_k^{a_{\rm s},b_{\rm s}}$. 
We obtain
\begin{equation}\label{eq_in_0}
n_{k,0}=
\max\left\{
\bigg\lfloor{}n_{k}\frac{m_{k,0}}{|{\mathcal Z}^{a_{\rm s},b_{\rm s}}_k|}-n_{k}
\Lambda(|{\mathcal Z}^{a_{\rm s},b_{\rm s}}_k|,n_k,\varepsilon''_{k,0}) \bigg\rfloor,0\right\},
\end{equation} 
except with error probability 
\begin{equation}\label{eps_0}
\varepsilon_{k,0}\leq \varepsilon'_{k,0}+\varepsilon''_{k,0},
\end{equation}
where the function $\Lambda(x,y,z)$ is defined as
$\Lambda(x,y,z)=\sqrt{(x-y+1)\ln{(z^{-1})}/(2xy)}$, 
and $\varepsilon'_{k,0}\leq \varepsilon_0+\sum_{a}\gamma_{a,b_{\rm s}}$.
 
\subsection{Estimation of $n_{k,1}$}\label{sub_n11}

To estimate $n_{k,1}$ we employ the same two-step method that we used to obtain $n_{k,0}$. That is, 
we first compute  
a lower bound for the number of indexes in ${\mathcal Z}^{a_{\rm s},b_{\rm s}}_k$ 
where both Alice and Bob sent a single-photon. We 
shall denote this quantity
as
$m_{k,1}$, which can be written as
\begin{equation}\label{mk11}
m_{k,1}={}
p_{a_{\rm s},b_{\rm s}|11,{\rm Z}}S_{k,11}-\Delta_{1},
\end{equation}
except with error probability $\varepsilon_{1}$, where the parameter
$\Delta_{1}=g(p_{a_{\rm s},b_{\rm s}|11,{\rm Z}}S_{k,11},\varepsilon_{1})$. 
Again, 
this statement can be proven with similar arguments to those used
to prove Eq.~(\ref{g5}).
Second, we calculate $n_{k,1}$ from $m_{k,1}$ using 
the Serfling inequality~[49]. 

According to Eq.~(\ref{mk11}), to compute $m_{k,1}$ 
we need to search
for a lower bound for $S_{k,11}$. This is what we do next. 
Our starting point is Eq.~(\ref{monday}). The estimation method is then divided into two steps. First, we 
cancel the terms ${\tilde S}_{k,0m}$ and ${\tilde S}_{k,n0}$ using Gaussian elimination. Second, we cancel either the parameter 
${\tilde S}_{k,12}$ or ${\tilde S}_{k,21}$, 
depending on the combination of intensities that are used in the first step; this will become clear below. 

Let us begin with the first step. For this, 
we introduce a vector of intensities
$v=[a_0,a_1,b_0,b_1]$ 
that satisfies
$a_0>a_1$ and $b_0>b_1$, with $a_i\in{\mathcal A}$ and
$b_i\in{\mathcal B}$. Then,
we find that the parameters 
$G_{k,v}$ defined below do not contain any term of the form ${\tilde S}_{k,0m}$ or ${\tilde S}_{k,n0}$,
with 
\begin{eqnarray}\label{cs_upp1}
G_{k,v}=|{\tilde {\mathcal Z}}^{a_0,b_0}_k|
+|{\tilde {\mathcal Z}}^{a_1,b_1}_k|-|{\tilde {\mathcal Z}}^{a_0,b_1}_k|-|{\tilde {\mathcal Z}}^{a_1,b_0}_k|. \quad\quad
\end{eqnarray}

Next, we move to the second step. Here,
we select another vector $v'=[a'_0,a'_1,b'_0,b'_1]$ that fulfils
the same conditions as $v$, and, moreover, satisfies the following constraints: 
$a_i=a'_j$, $a_{i\oplus1}>a'_{j\oplus1}$, 
$b_{i'}=b'_{j'}$ and $b_{i'\oplus1}>b'_{j'\oplus1}$
 for certain $i,j,i',j'=0,1$, and where the symbol $\oplus$ denotes the 
 modulo-$2$ addition. Then, we need to consider two cases. 
 
{\it Case $1$:} If $(a_0+a_1)/(a'_0+a'_1)>(b_0+b_1)/(b'_0+b'_1)$, we define the parameter $J_{k,v,v'}$ as
\begin{equation}\label{trn_station}
J_{k,v,v'}=
(b_0^2-b_1^2)(a_0-a_1)
G_{k,v'}-
(b_0^{\prime 2}-b_1^{\prime 2})(a'_0-a'_1)G_{k,v}. 
\end{equation}
Using Eqs.~(\ref{monday})-(\ref{cs_upp1}), we can rewrite $J_{k,v,v'}$ as
\begin{eqnarray}\label{dom_late_a}
J_{k,v,v'}&=&
\sum_{n,m=1}^\infty
\frac{c_{nm}}{n!m!}
{\tilde S}_{k,nm}
+ \delta_{k,v,v'},
\end{eqnarray}
where the coefficients $c_{nm}$ and $\delta_{k,v,v'}$ are given by
\begin{eqnarray}
c_{nm}&=&
(b_0^2-b_1^2)
(a_0-a_1)
(a_0^{\prime n}-a_1^{\prime n})
(b_0^{\prime m}-b_1^{\prime m})
-(b_0^{\prime 2}-b_1^{\prime 2})
(a'_0-a'_1)
(a_0^{n}-a_1^{n})
(b_0^{m}-b_1^{m}),  \\
\delta_{k,v,v'}&=&(b_0^2-b_1^2)(a_0-a_1)\delta_{k,v'}-
(b_0^{\prime 2}-b_1^{\prime 2})(a'_0-a'_1)\delta_{k,v}, \nonumber
\end{eqnarray}
and the parameters $\delta_k^{v}$ and $\delta_k^{v'}$ have the form
\begin{eqnarray}
\delta_k^{v}&=&{\tilde \delta}_{a_0,b_0}+{\tilde \delta}_{a_1,b_1}-{\tilde \delta}_{a_0,b_1}-{\tilde \delta}_{a_1,b_0}, \nonumber \\
\delta_k^{v'}&=&{\tilde \delta}_{a'_0,b'_0}+{\tilde \delta}_{a'_1,b'_1}-{\tilde \delta}_{a'_0,b'_1}-{\tilde \delta}_{a'_1,b'_0}.
\end{eqnarray}

From the conditions above, it is easy to show that $c_{11}\geq{}0$ and $c_{nm}\leq{}0$ when $n+m\geq3$.
 For simplicity, however, we de not include such proofs here. 
 Combining these results with Eq.~(\ref{dom_late_a}), we 
 obtain
\begin{eqnarray}\label{dom_late}
{\tilde S}_{k,11}\geq \frac{1}{c_{11}}\left(J_{k,v,v'}-\delta_{k,v,v'}\right).
\end{eqnarray}
Now, we need to compute an upper bound for $\delta_{k,v,v'}$. 
Using the fact that ${\tilde \delta}_{a,b}\in[-\Gamma_{a,b}, {\hat \Gamma}_{a,b}]$
except with 
error probability $\gamma_{a,b}$, we obtain that $\delta_{k,v,v'}\leq \Gamma_{k,v,v'}$ with
\begin{eqnarray}\label{ww2}
\Gamma_{k,v,v'}&=& 
(b_0^2-b_1^2)(a_0-a_1)
({\hat \Gamma}_{a'_0,b'_0}+{\hat \Gamma}_{a'_1,b'_1}+{\hat \Gamma}_{a'_0,b'_1}
+{\hat \Gamma}_{a'_1,b'_0})+
(b_0^{\prime 2}-b_1^{\prime 2})(a'_0-a'_1)
(\Gamma_{a_0,b_0}+\Gamma_{a_1,b_1} \nonumber \\
&+&\Gamma_{a_0,b_1}+\Gamma_{a_1,b_0}).
\end{eqnarray}

We find, therefore, that 
\begin{eqnarray}
S_{k,11}\geq s_{k,{\mathcal V}}=\max_{v,v'\in{\mathcal V}}\frac{\tau_{11}}{c_{11}}\left(J_{k,v,v'}-\Gamma_{k,v,v'}\right),
\end{eqnarray}
expect with error probability 
\begin{equation}\label{e_Z2}
\sum_{a,b} \gamma_{a,b},
\end{equation}
where ${\mathcal V}$ denotes the set of pairs of vectors $v$, $v'$, which satisfy the conditions
required in Case $1$. 

{\it Case $2$:} If $(a_0+a_1)/(a'_0+a'_1)\leq(b_0+b_1)/(b'_0+b'_1)$, we define the parameter $J_{k,v,v'}$ as
\begin{equation}\label{trn_station2}
J_{k,v,v'}=
(a_0^2-a_1^2)(b_0-b_1)
G^{v'}_k-
(a_0^{\prime 2}-a_1^{\prime 2})(b'_0-b'_1)G^{v}_k,
\end{equation}
and we proceed as in Case $1$. We
obtain
\begin{eqnarray}
S_{k,11}\geq s_{k,{\mathcal V'}}=\max_{v,v'\in{\mathcal V'}}\frac{\tau_{11}}{c_{11}}\left(J_{k,v,v'}-\Gamma_{k,v,v'}\right),
\end{eqnarray}
expect with error probability 
given by Eq.~(\ref{e_Z2}), 
where
${\mathcal V'}$ contains vectors $v$, $v'$, which satisfy the conditions
required in Case $2$. Now, the coefficient 
$c_{11}=(a_0-a_1)
(b_0-b_1)
(a'_0-a'_1)(b'_0-b'_1)
(a_0+a_1-a'_0-a'_1)$, and
\begin{eqnarray}
\Gamma_{k,v,v'}&=& 
(a_0^2-a_1^2)(b_0-b_1)
({\hat \Gamma}_{a'_0,b'_0}+{\hat \Gamma}_{a'_1,b'_1}+{\hat \Gamma}_{a'_0,b'_1}
+{\hat \Gamma}_{a'_1,b'_0})+
(a_0^{\prime 2}-a_1^{\prime 2})(b'_0-b'_1)
(\Gamma_{a_0,b_0}+\Gamma_{a_1,b_1} \nonumber \\
&+&\Gamma_{a_0,b_1}+\Gamma_{a_1,b_0}).
\end{eqnarray}

As a result, we obtain that $S_{k,11}$ is lower bounded by
\begin{equation} \label{Eqn:mZ11L}
S_{k,11}\geq\max\left\{s_{k,{\mathcal V}},s_{k,{\mathcal V'}},0\right\},
\end{equation}
except with error probability given by Eq.~(\ref{e_Z2}).

Finally, we use the Serfling inequality~[49] and find that
\begin{equation}\label{eq_ina}
n_{k,1}=
\max\left\{
\bigg\lfloor{}n_{k}\frac{m_{k,1}}{|{\mathcal Z}^{a_{\rm s},b_{\rm s}}_k|}
-n_{k}\Lambda(|{\mathcal Z}^{a_{\rm s},b_{\rm s}}_k|,n_k,\varepsilon''_{k,1}) \bigg\rfloor,0\right\},
\end{equation} 
except with error probability \cite{note_pa2}
\begin{equation}\label{eps_1}
\varepsilon_{k,1}\leq \varepsilon'_{k,1}+\varepsilon''_{k,1},
\end{equation}
where $\epsilon'_{k,1}\leq \varepsilon_1+\sum_{a,b} \gamma_{a,b}$.

\subsection{Estimation of $e_{k,1}$}\label{sub_ek1}

The procedure to estimate $e_{k,1}$ can be decomposed into three steps. First, 
we calculate a lower bound for the number of signals where Alice and Bob send a 
single-photon state
prepared in the basis $\rm X$, and where Charles declares the Bell state $k$. 
We will denote this quantity as ${\bar n}_{k,1}$. Second, we obtain an upper bound for 
the total number of errors in these signals.  We shall denote this parameter as
${\bar e}_{k,1}$. And, third, we use the Serfling result~[49] to compute
$e_{k,1}$ from $n_{k,1}$, ${\bar n}_{k,1}$ and ${\bar e}_{k,1}$. 

Suppose that we already completed the first two steps and we  
obtained
${\bar n}_{k,1}$ and ${\bar e}_{k,1}$.
Then, the number of signals where Alice and Bob send a single-photon state, 
and Charles declares the Bell state $k$, is
lower bounded by $n_{k,1}+{\bar n}_{k,1}$, 
with 
$n_{k,1}$ given by Eq.~(\ref{eq_ina}). Now, since these single-photon signals 
(when averaged over Alice's and Bob's key bit values) are basis independent, 
the Serfling inequality tells us that 
\begin{equation}\label{q1}
e_{k,1}=
\min\Bigg\{
\bigg\lceil n_{k,1}
\bigg(\frac{{\bar e}_{k,1}}{{\bar n}_{k,1}}\bigg)
+(n_{k,1}+{\bar n}_{k,1})\Upsilon(n_{k,1},{\bar n}_{k,1},\varepsilon'''_{k,e})\bigg\rceil,n_{k,1}\Bigg\},
\end{equation}
except with error probability 
\begin{equation}\label{eps_2}
\varepsilon_{k,e}\leq \varepsilon'_{k,e}+\varepsilon''_{k,e}+\varepsilon'''_{k,e},
\end{equation}
where the function $\Upsilon(x,y,y)$ is defined as
$\Upsilon(x,y,z)=\sqrt{(x+1)\ln{(z^{-1})}/(2y(x+y))}$.

Next, we calculate ${\bar n}_{k,1}$ and ${\bar e}_{k,1}$, together with their
associated error probabilities $\varepsilon'_{k,e}$ and $\varepsilon''_{k,e}$. 
To obtain ${\bar n}_{k,1}$
we use the same strategy presented in Appendix~\ref{sub_n11} to calculate 
a lower bound for $S_{k,11}$. 
We only need to replace
the basis $\rm Z$
with the basis $\rm X$ in all the mathematical expressions 
that appear in that Appendix. 
Thereby 
we find that ${\bar n}_{k,1}$ has a similar expression to that given by Eq.~(\ref{Eqn:mZ11L}), 
except with error probability
$\varepsilon'_{k,e}\leq\sum_{a,b}\gamma_{a,b}$.

Below we 
obtain ${\bar e}_{k,1}$. For this, 
we need, however, to introduce a new
group of index sets, whose elements we will denote by ${\mathcal E}_k^{a,b}$. These sets identify signals 
where Charles declared the Bell state $k$, Alice and Bob selected 
the intensity settings $a$ and $b$ and the basis $\rm X$, and, 
after applying the bit flip operation in the sifting step of the protocol, their bits differ. That is, 
$\{{\mathcal E}_k^{a,b}\}_{a,b}$ points to errors in the basis $\rm X$.

Also, let $E_{k,nm}$
denote the number of signals sent by 
Alice and Bob with $n$ and $m$ photons respectively, when they select 
the basis ${\rm X}$,  
Charles declares the 
Bell state $k$, and, after applying the bit flip operation in the sifting step, Alice's and Bob's bits differ. 
That is, ${\bar e}_{k,1}$ represents an upper bound for $E_{k,11}$. 

Our starting point is the size of the sets ${\mathcal E}_{k}^{a,b}$, i.e.,
\begin{equation}
|{\mathcal E}_{k}^{a,b}|=\sum_{n,m}p_{a,b|nm,{\rm X}}E_{k,nm}+\delta_{a,b}.
\end{equation}
This equation can be rewritten as
\begin{equation}
|{\tilde {\mathcal E}}^{a,b}_k|
=\sum_{n,m}\frac{a^nb^m}{n!m!}{\tilde E}_{k,nm}+{\tilde \delta}_{a,b},
\end{equation}
where $|{\tilde {\mathcal E}}^{a,b}_k|=e^{a+b}|{\mathcal E}^{a,b}_k|/p_{a,b,{\rm X}}$,  
${\tilde E}_{k,nm}=E_{k,nm}/\tau_{nm}$ with
$\tau_{nm}$ having the form of Eq.~(\ref{taoo}) but with $p_{a,b,{\rm X}}$ instead
of $p_{a,b,{\rm Z}}$, and
${\tilde \delta}_{a,b}=e^{a+b}\delta_{a,b}/p_{a,b,{\rm X}}$.

Next, we follow a similar procedure to that used in Appendix~\ref{sub_n11}. Now, however, 
we try to cancel out only
the terms ${\tilde E}_{k,0m}$ and ${\tilde E}_{k,n0}$. For this,
we introduce a vector
$v=[a_0,a_1,b_0,b_1]$ 
that satisfies
$a_0>a_1$ and $b_0>b_1$, with $a_i\in{\mathcal A}$ and
$b_i\in{\mathcal B}$, and 
we define the parameters 
$F_{k,v}$ as
\begin{eqnarray}
F_{k,v}&=&|{\tilde {\mathcal E}}^{a_0,b_0}_k|
+|{\tilde {\mathcal E}}^{a_1,b_1}_k|-|{\tilde {\mathcal E}}^{a_0,b_1}_k|-|{\tilde {\mathcal E}}^{a_1,b_0}_k|  
=(a_0-a_1)(b_0-b_1){\tilde E}_{k,11}+\delta_k^v \nonumber \\
&+& \sum_{\substack{n,m =1\\ (n,m)\neq(1,1)}}^{\infty}
\frac{(a_0^n-a_1^n)(b_0^m-b_1^m)}{n!m!}
{\tilde E}_{k,nm},\quad\quad  \label{wee}
\end{eqnarray}
where $\delta_k^{v}={\tilde \delta}_{a_0,b_0}+{\tilde \delta}_{a_1,b_1}-{\tilde \delta}_{a_0,b_1}-{\tilde \delta}_{a_1,b_0}$.
The third term on the r.h.s of Eq.~(\ref{wee}) is always greater or equal than zero. This means, in particular, 
that $F_{k,v}$ is lower bounded by
\begin{equation}\label{eqq}
F_{k,v}
\geq(a_0-a_1)(b_0-b_1){\tilde E}_{k,11}+\delta_k^v.
\end{equation}
Now, we need to compute a lower bound for $\delta_k^v$.
We have that 
each parameter  ${\tilde \delta}_{a,b}\in[-\Gamma_{a,b}, {\hat \Gamma}_{a,b}]$, 
with $\Gamma_{a,b}=e^{a+b}\Delta_{a,b}/p_{a,b,{\rm X}}$, 
${\hat \Gamma}_{a,b}=e^{a+b}{\hat \Delta}_{a,b}/p_{a,b,{\rm X}}$, 
$\Delta_{a,b}=g(|{\mathcal E}^{a,b}_k|,\varepsilon_{a,b}^4/16)$ and
${\hat \Delta}_{a,b}=g(|{\mathcal E}^{a,b}_k|,{\hat \varepsilon}_{a,b}^{3/2})$, except with 
error probability $\gamma_{a,b}=\epsilon_{a,b}+\varepsilon_{a,b}+{\hat \varepsilon}_{a,b}$. We obtain, therefore, that
\begin{equation}\label{eqq2}
\delta_k^v\geq
\Gamma_{k,v}=
-\Gamma_{a_0,b_0}-\Gamma_{a_1,b_1}-{\hat \Gamma}_{a_0,b_1}-{\hat \Gamma}_{a_1,b_0}.
\end{equation}

Finally, if we combine Eqs.~(\ref{eqq})-(\ref{eqq2}), 
we find that $E_{k,11}$ is upper bounded by
\begin{eqnarray} \label{off_stock}
E_{k,11}\leq\min_v\frac{(F_{k,v}-\Gamma_{k,v})\tau_{11}}{(a_0-a_1)(b_0-b_1)}={\bar e}_{k,1},
\end{eqnarray}
except with error probability $\varepsilon''_{k,e}\leq\sum_{a,b}\gamma_{a,b}$.

\section{Numerical estimation of 
$n_{k,0}$,
$n_{k,1}$ and $e_{k,1}$}\label{ap_numer}

In this Appendix we present 
a numerical method 
to calculate $n_{k,0}$,
$n_{k,1}$ and $e_{k,1}$ 
that is valid for any number of decoy states used by Alice and Bob, and
for any photon-number distribution of their signals. It may be used, for instance, with sources 
emitting phase-randomised weak coherent pulses, triggered spontaneous parametric down-conversion sources, 
and also with practical single-photon sources. More precisely, 
we
show that 
the
estimation of these parameters
can be 
written as
a linear program, 
which
 can be solved efficiently 
in polynomial time, and gives the optimum even for large dimensions~[43]. 

Let us introduce first some notations. In particular,  
let $N_{nm}$ denote the number of signals sent by Alice and Bob with $n$ and $m$ photons respectively,
when they select the 
${\rm Z}$ basis. And, let $N=\sum_{n,m} N_{nm}$ be the number of signals sent in the 
${\rm Z}$ basis. Using the Chernoff-Hoeffding inequality for i.i.d. random variables [34]-\cite{hoef}, we have that
\begin{equation}\label{brea}
N_{nm}= N(p_{nm|{\rm Z}}+\delta_{nm}),
\end{equation}
except with error probability $\gamma_{nm}=\varepsilon_{nm}+{\hat \varepsilon}_{nm}$.
The term 
$p_{nm|{\rm Z}}$ 
is a parameter that characterises the sources. It
represents the conditional probability that 
Alice and Bob send a signal with $n$ and $m$ photons respectively, 
given that they selected the basis ${\rm Z}$. The parameter
$\delta_{nm}$ 
lies in the interval $[-\Delta_{nm}, {\hat \Delta}_{nm}]$, with
\begin{eqnarray}
\Delta_{nm}&=&\min \left\{g(p_{nm|{\rm Z}}/N,\varepsilon_{nm}),p_{nm|{\rm Z}}\right\},  \label{st1} \\
{\hat \Delta}_{nm}&=&\min \left\{
k(N,p_{nm|{\rm Z}},{\hat \varepsilon}_{nm}), 1-p_{nm|{\rm Z}}\right\}. \label{st2} 
\end{eqnarray}
Here, the function
$k(x,y,z)=\ln{(z^{-2})}(1+\sqrt{1+4xy/\ln{(z^{-2})}})/2x$, and
the
second term 
on the r.h.s. of Eqs.~(\ref{st1})-(\ref{st2}) 
is due to the fact that
$N\geq N_{nm}$ $\forall n, m$. Also, we have that $\sum_{n,m} N(p_{nm|{\rm Z}}+\delta_{nm})$
is by definition equal to $N$, i.e., the terms $\delta_{nm}$
satisfy $\sum_{n,m}\delta_{nm}=0$. 

\subsection{Estimation of $n_{k,0}$}\label{ap1}

The procedure to calculate $n_{k,0}$ is similar to that used in the analytical approach presented 
in Appendix~\ref{sub_n11b}. 
First, we obtain the 
parameter $m_{k,0}$ and then we apply Eq.~(\ref{eq_in_0}). 

Next, we show that 
the search for $m_{k,0}$ can be formulated as a linear program. 
To do so, 
however, we need to reduce the number of unknown 
parameters $S_{k,nm}$ and
$\delta_{nm}$ to a finite set. 
For this, 
we first derive a lower and upper bound for the quantities
$|{\mathcal Z}^{a,b}_k|$. 

In particular, since $p_{a,b|nm,{\rm Z}}S_{k,nm}\geq{}0$ for all $n,m$,
from Eq.~(\ref{g5})
we have that
\begin{eqnarray}
|{\mathcal Z}^{a,b}_k|
\geq \sum_{n,m\in{\mathcal S}_{\rm cut}}p_{a,b|nm,{\rm Z}}S_{k,nm}+\delta_{a,b}.
\end{eqnarray}
Here, ${\mathcal S}_{\rm cut}$ denotes a finite set of indexes $n, m$, which
includes the case $n=m=1$. 
For instance, one may select 
${\mathcal S}_{\rm cut}=\{n,m \in {\mathbb N} \ {\rm with}\ n+m\leq{}M_{\rm cut} \}$,
for a prefixed value $M_{\rm cut}\geq{}2$. In this case, 
${\mathcal S}_{\rm cut}$ has $N_{\rm cut}=(M_{\rm cut}+1)(M_{\rm cut}+2)/2$ elements. 

Similarly, we also have that
\begin{eqnarray}\label{qwe}
\sum_{n,m\notin{\mathcal S}_{\rm cut}}p_{a,b|nm,{\rm Z}}S_{k,nm}&\leq&
\sum_{n,m\notin{\mathcal S}_{\rm cut}}p_{a,b|nm,{\rm Z}}N_{nm}\leq
\max_{kl\notin{\mathcal S}_{\rm cut}} p_{a,b|kl,{\rm Z}}
 \sum_{n,m\notin{\mathcal S}_{\rm cut}} 
N\big(p_{nm|{\rm Z}}+
\delta_{nm}\big)
\nonumber \\
&&=\max_{kl\notin{\mathcal S}_{\rm cut}} p_{a,b|kl,{\rm Z}}\ 
N\Bigg[1-\sum_{n,m\in{\mathcal S}_{\rm cut}}(p_{nm|{\rm Z}}+\delta_{nm})\Bigg].\quad\quad
\end{eqnarray}
In the first two inequalities of Eq.~(\ref{qwe}) we use $N_{nm}\geq S_{k,nm}\geq{}0$, 
together with 
Eq.~(\ref{brea}). The last equality uses 
$\sum_{n,m\notin{\mathcal S}_{\rm cut}}p_{nm|{\rm Z}}
=1-\sum_{n,m\in{\mathcal S}_{\rm cut}}p_{nm|{\rm Z}}$ and 
$\sum_{n,m\notin{\mathcal S}_{\rm cut}}\delta_{nm}
=-\sum_{n,m\in{\mathcal S}_{\rm cut}}\delta_{nm}$.
If we now combine Eqs.~(\ref{g5})-(\ref{qwe}) we obtain that 
\begin{equation}
|{\mathcal Z}^{a,b}_k|\leq\sum_{n,m\in{\mathcal S}_{\rm cut}}p_{a,b|nm,{\rm Z}}S_{k,nm}+\max_{kl\notin{\mathcal S}_{\rm cut}} p_{a,b|kl,{\rm Z}}
N\Bigg[1-\sum_{n,m\in{\mathcal S}_{\rm cut}}(p_{nm|{\rm Z}}+\delta_{nm})\Bigg]
+\delta_{a,b}. \quad\quad
\end{equation}

Moroever, using the Chernoff-Hoeffding inequality [34]-\cite{hoef},
it is straightforward to show that the term
$\sum_{n,m\in{\mathcal S}_{\rm cut}} \delta_{nm}$ lies in the interval 
$[-\Delta, {\hat \Delta}]$, with
\begin{eqnarray}
\Delta&=&\min \big\{g(p_{{\mathcal S}_{\rm cut}|{\rm Z}}/N,\varepsilon),p_{{\mathcal S}_{\rm cut}|{\rm Z}}\big\},  \label{st3} \\
{\hat \Delta}&=&\min \big\{
k(N,p_{{\mathcal S}_{\rm cut}|{\rm Z}},{\hat \varepsilon}), 1-p_{{\mathcal S}_{\rm cut}|{\rm Z}}\big\}, \label{st4} 
\end{eqnarray}
except with error probability $\gamma=\varepsilon+{\hat \varepsilon}$, 
where $p_{{\mathcal S}_{\rm cut}|{\rm Z}}=\sum_{n,m\in{\mathcal S}_{\rm cut}}p_{nm|{\rm Z}}$.

Based on the foregoing, we find that $m_{k,0}$ can be calculated 
using the following linear program
\begin{eqnarray}\label{sol_0}
\min \quad&&  \sum_{m=0}^{M_{\rm cut}}p_{a_{\rm s},b_{\rm s}|0m,{\rm Z}}S_{k,0m}
\nonumber \\
{\rm s. t.}\quad&& |{\mathcal Z}^{a,b}_k|
\geq \sum_{n,m\in{\mathcal S}_{\rm cut}}p_{a,b|nm,{\rm Z}}S_{k,nm}+\delta_{a,b}, \forall a,b \nonumber \\
&& |{\mathcal Z}^{a,b}_k|\leq\sum_{n,m\in{\mathcal S}_{\rm cut}}p_{a,b|nm,{\rm Z}}S_{k,nm}
+\max_{kl\notin{\mathcal S}_{\rm cut}} p_{a,b|kl,{\rm Z}}\
N\Bigg[1-\sum_{n,m\in{\mathcal S}_{\rm cut}}(p_{nm|{\rm Z}}+\delta_{nm})\Bigg]
+\delta_{a,b}, 
\forall a,b \nonumber \\
&& \sum_{a,b} \delta_{a,b}=0,  \quad{\hat \Delta}_{a,b}\geq \delta_{a,b} \geq -\Delta_{a,b}, \forall a,b\nonumber \\
&& N\left(p_{nm|{\rm Z}}+\delta_{nm}\right)\geq S_{k,nm} \geq 
0,
\forall n,m\in{\mathcal S}_{\rm cut} \nonumber \\
&& {\hat \Delta}_{nm}\geq \delta_{nm} \geq -\Delta_{nm}, \forall n,m\in{\mathcal S}_{\rm cut} \nonumber \\
&& {\hat \Delta}\geq{}\sum_{n,m\in{\mathcal S}_{\rm cut}}
\delta_{nm}
 \geq -\Delta. 
\end{eqnarray}
The constraint 
$\sum_{a,b} \delta_{a,b}=0$ is due to the fact that 
$\sum_{a,b}|{\mathcal Z}^{a,b}_k|$ is by definition 
equal to $\sum_{n,m}S_{k,nm}$. 

The linear program 
given by Eq.~(\ref{sol_0})
has $2N_{\rm cut}+(d_{\rm A}+1)(d_{\rm B}+1)$ unknown
parameters $S_{k,nm}$, $\delta_{nm}$ and
$\delta_{a,b}$. Here,
$d_{\rm A}$ ($d_{\rm B}$) denotes the number of decoy intensity settings used 
by Alice (Bob). 
The number of known elements
is $(d_{\rm A}+1)(d_{\rm B}+1)N_{\rm cut}+(d_{\rm A}+1)(d_{\rm B}+1)+1+N_{\rm cut}$. 
These are the 
terms  
$p_{a,b|nm,{\rm Z}}$, $|{\mathcal Z}^{a,b}_k|$, $\max_{kl\notin{\mathcal S}_{\rm cut}} p_{a,b|kl,{\rm Z}}$ and
$p_{nm|{\rm Z}}$. Finally, given the tolerated failure probabilities
$\varepsilon_{a,b}$, ${\hat \varepsilon}_{a,b}$, $\varepsilon_{nm}$, ${\hat \varepsilon}_{nm}$,
$\varepsilon$ and ${\hat \varepsilon}$, the value of 
${\hat \Delta}_{a,b}$, $\Delta_{a,b}$, ${\hat \Delta}_{nm}$, $\Delta_{nm}$, ${\hat \Delta}$ and $\Delta$
is also known. 

If we denote the solution to the optimisation problem 
given by Eq.~(\ref{sol_0})
as $n_{\rm sol}$, then 
from Eq.~(\ref{mk0})
we have that
\begin{equation}\label{nw3}
m_{k,0}=\max\big\{\lfloor n_{\rm sol}-g(n_{\rm sol},\varepsilon_{0})\rfloor,0\big\},
\end{equation}
except with error probability $\varepsilon'_{k,0}$ given by
\begin{eqnarray}
&&\varepsilon'_{k,0}\leq
\varepsilon_{0}+
\gamma+\sum_{a,b}\gamma_{a,b}+\sum_{n,m\in{\mathcal S}_{\rm cut}}\gamma_{nm}.
\end{eqnarray}

\subsection{Estimation of $n_{k,1}$}\label{ap1b}

To obtain $n_{k,1}$ we use again the same two-step technique introduced in 
Appendix~\ref{sub_n11}. That is, we first
calculate $m_{k,1}$, and then we use Eq.~(\ref{eq_ina}). To estimate $m_{k,1}$, 
we first obtain a lower bound for 
$S_{k,11}$, and then we apply Eq.~(\ref{mk11}). 
In so doing, we reduce the problem of calculating
$n_{k,1}$ to that of finding a lower bound for 
$S_{k,11}$. This is what we do below. 

For this, we reuse the linear program given by Eq.~(\ref{sol_0}), 
only substituting its 
linear objective function 
with $S_{k,11}$.
If we denote the solution to this program as $n_{\rm sol}$, then 
from Eq.~(\ref{mk11})
we have that
\begin{equation}\label{nw4b}
m_{k,1}=\max\big\{\lfloor p_{a_{\rm s},b_{\rm s}|11,{\rm Z}}
n_{\rm sol}-g(p_{a_{\rm s},b_{\rm s}|11,{\rm Z}}n_{\rm sol},\varepsilon_{1})\rfloor,0\big\},\quad\quad
\end{equation}
except with error probability $\varepsilon'_{k,1}$ given by
\begin{eqnarray}
&&\varepsilon'_{k,1}\leq
\varepsilon_{1}+
\gamma+\sum_{a,b}\gamma_{a,b}+\sum_{n,m\in{\mathcal S}_{\rm cut}}\gamma_{nm}.
\end{eqnarray}

\subsection{Estimation of $e_{k,1}$}\label{ap1c}

Again, to estimate $e_{k,1}$ we follow the same steps introduced 
in Appendix~\ref{sub_ek1}. That is, we calculate the parameters ${\bar n}_{k,1}$ and 
${\bar e}_{k,1}$, and then we apply Eq.~(\ref{q1}). 

\subsubsection{Estimation of ${\bar n}_{k,1}$}

To obtain ${\bar n}_{k,1}$ we once more reuse the linear program given by Eq.~(\ref{sol_0}), only making 
the following three changes. First, all the parameters now refer 
to the $\rm X$ basis rather than the $\rm Z$ basis. For example, 
$S_{k,nm}$ will denote the number of signals sent by Alice and Bob with
$n$ and $m$ photons respectively, when they select the $\rm X$ basis and 
Charles announces the Bell state $k$. And, likewise for the other parameters.  
Second, we substitute the probabilities $p_{a,b|nm,{\rm Z}}$ and $p_{nm|{\rm Z}}$ 
with $p_{a,b|nm,{\rm X}}$ and $p_{nm|{\rm X}}$ respectively, and the sets 
${\mathcal Z}_k^{a,b}$ with ${\mathcal X}_k^{a,b}$. Third, we replace the 
linear objective function 
by $S_{k,11}$. 

Then, if $n_{\rm sol}$ denotes the solution to this program, 
we have that
\begin{equation}\label{nw4}
{\bar n}_{k,1}=\max\big\{\lfloor n_{\rm sol}\rfloor,0\big\},
\end{equation}
except with error probability $\varepsilon'_{k,e}$ given by
\begin{eqnarray}
&&\varepsilon'_{k,e}\leq
\gamma+\sum_{a,b}\gamma_{a,b}+\sum_{n,m\in{\mathcal S}_{\rm cut}}\gamma_{nm}.
\end{eqnarray}

\subsubsection{Estimation of ${\bar e}_{k,1}$}

By
using the same line of reasoning as in the previous sections, it is easy to show that 
${\bar e}_{k,1}$ can be calculated with the following 
linear program
\begin{eqnarray}\label{e_11}
\max \quad&&  E_{k,11}
\nonumber \\
{\rm s. t.}\quad&& |{\mathcal E}^{a,b}_k|
\geq \sum_{n,m\in{\mathcal S}_{\rm cut}}p_{a,b|nm,{\rm X}}E_{k,nm}+\delta_{a,b}, \forall a,b \nonumber \\
&& |{\mathcal E}^{a,b}_k|\leq\sum_{n,m\in{\mathcal S}_{\rm cut}}p_{a,b|nm,{\rm X}}E_{k,nm}
+\max_{kl\notin{\mathcal S}_{\rm cut}} p_{a,b|kl,{\rm X}}\
N\Bigg[1-\sum_{n,m\in{\mathcal S}_{\rm cut}}(p_{nm|{\rm X}}+\delta_{nm})\Bigg]
+\delta_{a,b}, 
\forall a,b \nonumber \\
&& \sum_{a,b} \delta_{a,b}=0,  \quad{\hat \Delta}_{a,b}\geq \delta_{a,b} \geq -\Delta_{a,b}, \forall a,b\nonumber \\
&& N\left(p_{nm|{\rm X}}+\delta_{nm}\right)\geq E_{k,nm} \geq 
0,
\forall n,m\in{\mathcal S}_{\rm cut} \nonumber \\
&& {\hat \Delta}_{nm}\geq \delta_{nm} \geq -\Delta_{nm}, \forall n,m\in{\mathcal S}_{\rm cut} \nonumber \\
&& {\hat \Delta}\geq{}\sum_{n,m\in{\mathcal S}_{\rm cut}}
\delta_{nm}
 \geq -\Delta,
\end{eqnarray}
where the definition of the different parameters is analogous to that of the previous sections, only
substituting ${\mathcal Z}^{a,b}_k$ with ${\mathcal E}^{a,b}_k$, $p_{nm|{\rm Z}}$ with $p_{nm|{\rm X}}$, 
$p_{{\mathcal S}_{\rm cut}|{\rm Z}}$ with $p_{{\mathcal S}_{\rm cut}|{\rm X}}$, and $N$ with the number of signals sent by Alice and Bob
in the $\rm X$ basis. If $n_{\rm sol}$ denotes the solution to this program then 
\begin{equation}\label{ooo}
{\bar e}_{k,1}=\min\{\lceil n_{\rm sol}\rceil,{\bar n}_{k,1}\},
\end{equation}
except with error probability $\varepsilon''_{k,e}$ given by
\begin{eqnarray}
&&\varepsilon''_{k,e}\leq
\gamma+\sum_{a,b}\gamma_{a,b}+\sum_{n,m\in{\mathcal S}_{\rm cut}}\gamma_{nm}.
\end{eqnarray}

\section{Chernoff bound}\label{chernoff}

Here we present the proof for Claim $1$ introduced in Appendix~\ref{sketch}. Also, we  
demonstrate a generalised version of it that can be applied when the conditions required in the 
Claim are not fulfilled. For simplicity, 
we divide this section into three parts. First, we introduce and demonstrate Claim $2$ below, which assumes that
the mean value $\mu$ is known. Second, we use this result to prove Claim $1$, which 
considers that $\mu$ is unknown. 
Third, we present a generalised version of Claim $1$ (see Claim $3$ below)
that can be employed when $(2\varepsilon^{-1})^{1/\mu_{\rm L}}>\exp{\left[3/(4\sqrt{2})\right]^2}$ and/or 
$({\hat \varepsilon}^{-1})^{1/\mu_{\rm L}}\geq\exp{(1/3)}$. 

\noindent{\bf Claim 2.} {\it Let 
$X_1, X_2, \ldots, X_n$, be a set of independent Bernoulli random variables that 
satisfy ${\rm Pr}(X_i=1)=p_i$, and let $X=\sum_{i=1}^n X_i$ and
$\mu=E[X]=\sum_{i=1}^n p_i$, 
where $E[\cdot]$ denotes the mean value. 
Let $x$ be the observed outcome of $X$ for a given trial (i.e., $x\in{\mathbb N}^+$).
When $(2\varepsilon^{-1})^{1/\mu}\leq\exp{\left[3/(4\sqrt{2})\right]^2}$ and
$({\hat \varepsilon}^{-1})^{1/\mu}<\exp{(1/3)}$ 
for certain $\varepsilon, {\hat \varepsilon}> 0$, we have that 
$x$ satisfies
\begin{equation}
x=\mu+\delta,
\end{equation}
except with error probability $\gamma=\varepsilon+{\hat \varepsilon}$, where the 
parameter $\delta\in[-\Delta,{\hat \Delta}]$, with 
$\Delta=g(x, \varepsilon^4/16)$,
${\hat \Delta}=g(x, {\hat \varepsilon}^{3/2})$ and $g(x,y)=\sqrt{2x\ln{(y^{-1}})}$.
Here $\varepsilon$ (${\hat \varepsilon}$) denotes the 
probability that $x<\mu-\Delta$ ($x>\mu+{\hat \Delta}$).
}

That is, Claim $2$ implies that the observed outcome $x$ of $X$ for a given trial satisfies
\begin{equation}
\mu+\sqrt{2x\ln{( {\hat \varepsilon}^{-{3/2}}})}\geq x\geq \mu-\sqrt{2x\ln{(16\varepsilon^{-4}})},
\end{equation}
except with error probability $\gamma=\varepsilon+{\hat \varepsilon}$,
given that both $(2\varepsilon^{-1})^{1/\mu}\leq\exp{\left[3/(4\sqrt{2})\right]^2}$ and
$({\hat \varepsilon}^{-1})^{1/\mu}<\exp{(1/3)}$. To simplify the notation in the proof below, 
we will denote these last two conditions as $C_1$ and $C_2$, respectively. 

{\it Proof.} Claim $2$ can be equivalently written as
\begin{equation}
{\rm Pr}\big[
\mu+\sqrt{2X\ln{({\hat \varepsilon}^{-3/2}})}
\geq X\geq\mu-\sqrt{2X\ln{(16\varepsilon^{-4})}}
|\
C_1 \land
C_2
\big]\geq 1-\varepsilon-{\hat \varepsilon}.
\end{equation}
To prove this statement, it is sufficient to show that
\begin{eqnarray} \label{eqa}
&&{\rm Pr}\left[X-\mu>\sqrt{2X\ln{({\hat \varepsilon}^{-3/2})}}\ |\ C_2\right]\leq {\hat \varepsilon}, \label{eqa1} \\
&&{\rm Pr}\left[\mu-X>\sqrt{2X\ln{(16\varepsilon^{-4})}}\ |\ C_1\right]\leq \varepsilon. \label{eqa2}
\end{eqnarray}

Let us begin by demonstrating Eq.~(\ref{eqa1}). 
Our starting point is a multiplicative form of the Chernoff 
bound for independent random variables\cite{bound1,bound2,bound3}. In particular, we have that
\begin{eqnarray}\label{eq_cher} 
{\rm Pr}\left[X>(1+\epsilon)\mu\right]&\leq& e^{-\frac{\mu\epsilon^2}{3}},
\end{eqnarray}
for $0<\epsilon<1$. 
This equation can be equivalently written as
\begin{eqnarray} \label{eq_cherc} 
{\rm Pr}\left[X-\mu>\sqrt{2\mu\ln{(\xi^{-3/2})}}\right]&\leq& \xi, 
\end{eqnarray}
for $(\xi^{-1})^{1/\mu}<\exp{(1/3)}$.
Now, to prove the first equation in Eq.~(\ref{eqa}) we consider two cases: 
$X\leq \mu$ and $X > \mu$. More precisely, we have that
\begin{eqnarray}
&&{\rm Pr}\big[X-\mu>\sqrt{2X\ln{({\hat \varepsilon}^{-3/2})}}\land X\leq\mu\ 
|\ C_2\big]=0,\quad
\end{eqnarray}
since both events are mutually exclusive. For the second case, we have that
\begin{eqnarray}\label{unobb}
{\rm Pr}\big[X-\mu>\sqrt{2X\ln{({\hat \varepsilon}^{-3/2})}}\land X>\mu\ |\ C_2\big] 
&\leq&{\rm Pr}\big[X-\mu>\sqrt{2\mu\ln{({\hat \varepsilon}^{-3/2})}} \land X>\mu\ 
|\ C_2\big] \nonumber \\
&\leq&{\rm Pr}\big[X-\mu>\sqrt{2\mu\ln{({\hat \varepsilon}^{-3/2})}}\ |\ C_2\big]
\leq{\hat \varepsilon},
\end{eqnarray}
where in the first inequality we have used the fact that $\sqrt{2X\ln{({\hat \varepsilon}^{-3/2})}}>\sqrt{2\mu\ln{({\hat \varepsilon}^{-3/2})}}$
when $X>\mu$, in the second inequality we have used ${\rm Pr}[A \land B|C]\leq{\rm Pr}[A|C]$,
and in the
last inequality we have used Eq.~(\ref{eq_cherc}). 

Let us now prove Eq.~(\ref{eqa2}). 
Again, our starting point is a multiplicative form of the Chernoff 
bound for independent random variables\cite{bound1,bound2,bound3}. More precisely, 
\begin{eqnarray}
{\rm Pr}\left[X<(1-\epsilon)\mu\right]&\leq& e^{-\frac{\mu\epsilon^2}{2}}, \label{eq_cher2}
\end{eqnarray}
with $0<\epsilon<1$. This statement can be rewritten as
\begin{eqnarray}
{\rm Pr}\left[\mu-X>\sqrt{2\mu\ln{(2\xi^{-1})}}\right]&\leq& \xi/2, \label{eq_cher22}
\end{eqnarray}
for $(2\xi^{-1})^{1/\mu}<\exp{(1/2)}$. That is, Eq.~(\ref{eq_cher22}) is also valid when 
$(2\xi^{-1})^{1/\mu}\leq\exp{[3/(4\sqrt{2})]^2}$ since 
$\exp{[3/(4\sqrt{2})]^2}<\exp{(1/2)}$.
Now, we evaluate three cases:
$X\geq\mu$, $\mu/k\leq X<\mu$ and $X<\mu/k$ with $4\geq k>1$. The first case corresponds to
\begin{eqnarray}
{\rm Pr}\left[\mu-X>\sqrt{2X\ln{(16\varepsilon^{-4})}} \land X\geq\mu\ |\ C_1\right]=0, \quad
\end{eqnarray}
since both events are mutually exclusive. Let us now consider the 
second case, i.e.,
\begin{eqnarray}
{\rm Pr}\left[\mu-X>\sqrt{2X\ln{(16\varepsilon^{-4})}} \land \mu/k\leq X<\mu\ |\ C_1\right]
&\leq& {\rm Pr}\left[\mu-X>\frac{2}{\sqrt{k}}\sqrt{2\mu\ln{(2\varepsilon^{-1})}} \land \mu/k\leq X<\mu\ |\ C_1\right] \nonumber \\
&\leq&{\rm Pr}\left[\mu-X>\sqrt{2\mu\ln{(2\varepsilon^{-1})}} \land \mu/k\leq X<\mu\ |\ C_1\right] \nonumber \\
&\leq& 
{\rm Pr}\left[\mu-X>\sqrt{2\mu\ln{(2\varepsilon^{-1})}}\ |\ C_1\right]\leq
\varepsilon/2.
\end{eqnarray} 
In the first inequality we have used the fact that 
$X\geq\mu/k$ and $\ln{(16\varepsilon^{-4})}=4\ln{(2\varepsilon^{-1})}$, 
in the second inequality we have used $2/\sqrt{k}\geq1$ when 
$4\geq k>1$, in the third inequality we have used again
${\rm Pr}[A \land B|C]\leq {\rm Pr}[A|C]$, and
 in the last inequality we have used Eq.~(\ref{eq_cher22}).
Let us now consider the third case, i.e., 
\begin{equation}\label{eq1}
{\rm Pr}\left[\mu-X>\sqrt{2X\ln{(16\varepsilon^{-4})}} \land X<\mu/k\ |\ C_1\right]
\leq {\rm Pr}\left[X<\mu/k\ |\ C_1]={\rm Pr}[\mu-X>\chi\mu\ |\ C_1\right],
\end{equation}
with $\chi=(k-1)/k$. When $(2\varepsilon^{-1})^{1/\mu}\leq\exp{\left[(k-1)/(\sqrt{2}k)\right]^2}$ 
(i.e, when $\chi\mu\geq\sqrt{2\mu\ln{(2\varepsilon^{-1})}}$)
we have that
\begin{equation}\label{eq2}
{\rm Pr}\left[\mu-X>\chi\mu\ |\ (2\varepsilon^{-1})^{1/\mu}\leq
e^{\left[(k-1)/(\sqrt{2}k)\right]^2}\right] 
\leq{\rm Pr}\bigg[\mu-X>\sqrt{2\mu\ln{(2\varepsilon^{-1})}} 
 |\ (2\varepsilon^{-1})^{1/\mu}\leq e^{\left[(k-1)/(\sqrt{2}k)\right]^2}\bigg]\leq \varepsilon/2,
\end{equation}
where in the last inequality we have used Eq.~(\ref{eq_cher22}). That is, if we select 
$k=4$, then from Eqs.~(\ref{eq1})-(\ref{eq2}) we have that
\begin{equation}
{\rm Pr}\left[\mu-X>\sqrt{2X\ln{(16\varepsilon^{-4})}} \land X<\mu/4\ |\ C_1\right]\leq \varepsilon/2.
\end{equation}

Combining the results above we find that
\begin{eqnarray}
&&{\rm Pr}\Big[\mu-X>\sqrt{2X\ln{(16\varepsilon^{-4})}}\ |\ C_1\Big]
\leq \varepsilon. \quad \blacksquare
\end{eqnarray}

Next, we will use the Claim $2$ above to prove the Claim $1$ introduced in Appendix~\ref{sketch}.
For this, we only need to derive a lower bound for the mean value $\mu$, which we will 
denote as $\mu_{\rm L}$, as a function of 
the observed outcome $x$ of $X$.
This can be done 
using the Hoeffding inequality~\cite{hoef}. It states that 
\begin{equation}
{\rm Pr}\left[\mu\leq X-t\right]\leq e^{-\frac{2t^2}{n}},
\end{equation}
for $t>0$.
This condition can be equivalently written as
\begin{equation}
{\rm Pr}\left[\mu\leq X-\sqrt{n/2\ln{(1/\epsilon)}}\right]\leq\epsilon
\end{equation}
That is, with probability $1-\epsilon$ we have that
\begin{equation}
\mu>\mu_{\rm L}=x-\sqrt{n/2\ln{(1/\epsilon)}}.
\end{equation}
Then, if 
$(2\varepsilon^{-1})^{1/\mu_{\rm L}}\leq\exp{\left[3/(4\sqrt{2})\right]^2}$ and 
$({\hat \varepsilon}^{-1})^{1/\mu_{\rm L}}<\exp{(1/3)}$
we have that the conditions $(2\varepsilon^{-1})^{1/\mu}\leq\exp{\left[3/(4\sqrt{2})\right]^2}$ and
$({\hat \varepsilon}^{-1})^{1/\mu}<\exp{(1/3)}$ are also satisfied except with probability $\epsilon$. 
This is so because $(2\varepsilon^{-1})^{1/\mu}\leq(2\varepsilon^{-1})^{1/\mu_{\rm L}}$ and
 $({\hat \varepsilon}^{-1})^{1/\mu}\leq({\hat \varepsilon}^{-1})^{1/\mu_{\rm L}}$.
If the 
random variable $X$ satisfies the two conditions above, then from Claim $2$ we have that any observed outcome
${\bar x}$ of $X$ can be written as
${\bar x}=\mu+\delta$,
except with error probability $\gamma=\varepsilon+{\hat \varepsilon}$, where the 
parameter $\delta\in[-\Delta,{\hat \Delta}]$, with 
$\Delta=g({\bar x}, \varepsilon^4/16)$ and
${\hat \Delta}=g({\bar x}, {\hat \varepsilon}^{3/2})$. Since this result applies to any observed outcome
${\bar x}$ of $X$ it applies, in particular, to the 
outcome $x$. This concludes the proof of Claim $1$.  

To finish this section, we introduce now a generalised version of Claim $1$
(see Claim $3$ below). It
can be applied when the conditions $(2\varepsilon^{-1})^{1/\mu_{\rm L}}\leq\exp{\left[3/(4\sqrt{2})\right]^2}$ and/or 
$({\hat \varepsilon}^{-1})^{1/\mu_{\rm L}}<\exp{(1/3)}$ are not fulfilled. 

\noindent{\bf Claim 3.} {\it Let 
$X_1, X_2, \ldots, X_n$, be a set of independent Bernoulli random variables that 
satisfy ${\rm Pr}(X_i=1)=p_i$, and let $X=\sum_{i=1}^n X_i$ and
$\mu=E[X]=\sum_{i=1}^n p_i$, 
where $E[\cdot]$ denotes the mean value. Let
$x$ be 
the observed outcome of $X$ for a given trial (i.e., $x\in{\mathbb N}^+$)
and $\mu_{\rm L}=x-\sqrt{n/2\ln{(1/\epsilon)}}$ for 
certain $\epsilon>0$. Then, 
we have that 
$x$ satisfies
\begin{equation}\label{cher_met2}
x=\mu+\delta,
\end{equation}
except with error probability $\gamma$, where the 
parameter $\delta\in[-\Delta,{\hat \Delta}]$. 
Let $test_1$, $test_2$ and $test_3$ denote, respectively, 
the following three conditions:   
$(2\varepsilon^{-1})^{1/\mu_{\rm L}}\leq\exp{\left[3/(4\sqrt{2})\right]^2}$,
$({\hat \varepsilon}^{-1})^{1/\mu_{\rm L}}<\exp{(1/3)}$  and
$({\hat \varepsilon}^{-1})^{1/\mu_{\rm L}}<\exp{[(2e-1)/2]^2}$ for 
certain $\varepsilon, {\hat \varepsilon}>0$, and 
let $g(x,y)=\sqrt{2x\ln{(y^{-1}})}$. Now:
\begin{enumerate}
\item When $test_1$ and $test_2$ are fulfilled, we have that
$\gamma=\epsilon+\varepsilon+{\hat \varepsilon}$, 
$\Delta=g(x, \varepsilon^4/16)$ and
${\hat \Delta}=g(x, {\hat \varepsilon}^{3/2})$. 
\item When $test_1$ and $test_3$ are fulfilled (and $test_2$ is not fulfilled), we have that 
$\gamma=\epsilon+\varepsilon+{\hat \varepsilon}$, 
$\Delta=g(x, \varepsilon^4/16)$ and
${\hat \Delta}=g(x, {\hat \varepsilon}^{2})$. 
\item When $test_1$ is fulfilled and $test_3$ is not fulfilled, we have that $\gamma=\epsilon+\varepsilon+{\hat \varepsilon}$, 
$\Delta=g(x, \varepsilon^4/16)$ and ${\hat \Delta}=\sqrt{(n/2)\log{(1/\varepsilon)}}$.
\item When 
$test_1$ is not fulfilled and $test_2$ is fulfilled, we have that
$\gamma=\epsilon+\varepsilon+{\hat \varepsilon}$, 
$\Delta=\sqrt{(n/2)\log{(1/\varepsilon)}}$ and
${\hat \Delta}=g(x, {\hat \varepsilon}^{3/2})$.
\item When $test_1$ and $test_2$ are not fulfilled, and $test_3$ is fulfilled, we have that
$\gamma=\epsilon+\varepsilon+{\hat \varepsilon}$, 
$\Delta=\sqrt{(n/2)\log{(1/\varepsilon)}}$ and
${\hat \Delta}=g(x, {\hat \varepsilon}^{2})$.
\item When $test_1$, $test_2$ and $test_3$ are not fulfilled, we have that
$\gamma=\varepsilon+{\hat \varepsilon}$, 
$\Delta={\hat \Delta}=\sqrt{(n/2)\log{(1/\varepsilon)}}$. 
\end{enumerate}
}

{\it Proof.} Item $1$ is the Claim $1$. The proof for item $2$ is basically the same as 
that used to prove item $1$, only substituting Eq.~(\ref{eq_cher}) by
\begin{eqnarray}\label{lst}
{\rm Pr}\left[X>(1+\epsilon)\mu\right]&\leq& e^{-\frac{\mu\epsilon^2}{4}},
\end{eqnarray}
for $0<\epsilon<2e-1$.
For item $3$ we combine the proof of item $1$ for the lower tail with the Hoeffding inequality for the 
upper tail~\cite{hoef}. 
Item $4$ combines 
the proof of item $1$ for the upper tail with the Hoeffding inequality for the 
lower tail. Item $5$ combines the proof of item $2$ for the upper tail 
with the Hoeffding inequality for the 
lower tail. Finally, item $6$ uses the Hoeffding inequality for both the upper and lower tails. 


\bibliographystyle{apsrev}

\end{document}